\documentclass[12pt,amsmath,amssymb,aps]{revtex4-1}
\pdfoutput=1
\usepackage{color} 
\usepackage{slashed}
\usepackage{graphicx}

\def\beq{\begin{equation}} \def\eeq{\end{equation}}
\def\beqn{\begin{eqnarray}} \def\eeqn{\end{eqnarray}}
\def\bom#1{{\mbox{\boldmath $#1$}}} \def\to{\rightarrow}
\def\nn{\nonumber} 

\def\li#1{\mathrm{Li_2}\left(#1\right)}
\def\ln#1{\mathrm{log}\left(#1\right)}
\def\Eq#1{Eq.~(\ref{#1})}

\newcommand\KerLO{\la {\cal{P}}^{(0)}_{q_1 \gamma_2} \ra}
\newcommand\KerLOA{\la {\cal{P}}^{(0)}_{q_1 \bar{q}_2 \gamma_3} \left.\right|_{\epsilon^0} \ra}

\newcommand\IDC{\textbf{Id}}
\newcommand\SUNT{\textit{\textbf{T}}}

\newcommand\IC{\textbf{\textit{I}}}

\newcommand\Spamplitude{{\rm Sp}}

\newcommand\CG{c_{\Gamma}}

\newcommand\as{\alpha_{\mathrm{S}}}
\newcommand\g{g_{\mathrm{S}}}

\def\ep{\epsilon}

\def\wp{\widetilde P}

\def\beq{\begin{equation}}
\def\eeq{\end{equation}}
\def\beeq{\begin{eqnarray}}
\def\eeeq{\end{eqnarray}}

\def\bom#1{{\mbox{\boldmath $#1$}}}
\def\to{\rightarrow}

\newcommand{\la}{\langle}
\newcommand{\ra}{\rangle}

\def\nn{\nonumber}

\def\ID{1 \kern -.45 em 1}
\def\RS{{\scriptscriptstyle\rm R\!.S\!.}}
\def\cdr{{\scriptscriptstyle\rm C\!.D\!.R\!.}}
\def\dimr{{\scriptscriptstyle\rm D\!.R\!.}}

\def\sp{{\bom{Sp}}}
\def\ket#1{|{#1}\ra}

\def\cbet0{b_0}

\begin{document}

\begin{flushright}
LPN14-099\\ IFIC/14-54
\end{flushright}

\title{Triple collinear splitting functions at NLO for scattering processes with photons}

\author{Germ\'an F. R. Sborlini~$^{(a, b)}$} \email{gfsborlini@df.uba.ar}
\author{Daniel de Florian~$^{(a)}$} \email{deflo@df.uba.ar}  
\author{Germ\'an Rodrigo~$^{(b)}$}  \email{german.rodrigo@csic.es}
\affiliation{
${}^{(a)}$Departamento de F\'\i sica and IFIBA, FCEyN, Universidad de Buenos Aires, 
(1428) Pabell\'on 1 Ciudad Universitaria, Capital Federal, Argentina \\
${}^{(b)}$Instituto de F\'{\i}sica Corpuscular, 
Universitat de Val\`encia - Consejo Superior de Investigaciones Cient\'{\i}ficas,
Parc Cient\'{\i}fic, E-46980 Paterna, Valencia, Spain}

\date{\today}

\begin{abstract}
We present splitting functions in the triple collinear limit at 
next-to-leading order. The computation was performed in the context 
of massless QCD+QED, considering only processes which include at least one photon. 
Through the comparison of the IR divergent structure of splitting amplitudes 
with the expected known behavior, we were able to check our results. 
Besides that we implemented some consistency checks based on symmetry 
arguments and cross-checked the results among them. 
Studying photon-started processes, we obtained very compact results.
\end{abstract}

\maketitle

\section{Introduction}

Studying the collinear limit of scattering amplitudes is a key point for improving 
theoretical computations. This is a requirement to compare theoretical predictions with the highly-accurate results provided by the LHC and other experiments at colliders. 
In this context, understanding the singular behavior of scattering amplitudes in 
the multiple-collinear limit is necessary to obtain higher order QCD corrections 
to several processes. 

It is a well known fact that, in certain general kinematical configurations, 
strict collinear factorization is fulfilled~\cite{Collins:1989gx, Catani:2011st}. 
These factorization properties establish that it is possible to isolate the singular 
behavior of scattering amplitudes into universal factors called splitting 
amplitudes~\cite{Berends:1987me, Mangano:1990by}. 
At the level of squared amplitudes, the Altarelli-Parisi (AP) kernels or 
splitting functions control the infrared (IR) singular behavior. 
Moreover, splitting functions govern the evolution of parton distributions 
functions (PDF) and fragmentation functions (FF) through 
 the DGLAP equation~\cite{Altarelli:1977zs}. 
Also, they are the main ingredient in parton-shower generators and they are crucial 
for subtraction methods~\cite{Catani:1996vz} to compute 
physical cross sections at higher orders in perturbation theory. 

Double-collinear splitting amplitudes and squared amplitudes have been computed at 
one-loop~\cite{Bern:1998sc, Bern:1999ry, Bern:1993qk, Bern:1994zx,Bern:1995ix,Kosower:1999rx} 
and two-loop level~\cite{Bern:2004cz, Badger:2004uk, Vogt:2005dw, Vogt:2004mw, Moch:2004pa}. 
Splitting kernels beyond the double collinear limit, however, are known at the 
tree-level only~\cite{Campbell:1997hg, Catani:1998nv, Catani:1999ss, DelDuca:1999ha, Birthwright:2005ak, Birthwright:2005vi},
although some specific results for the triple collinear limit of one-loop amplitudes 
are already available~\cite{Catani:2003vu}. 

In this article, we present explicit results for unpolarized squared splitting 
amplitudes at next-to-leading order (NLO) in QCD
for triple collinear processes involving at least one photon.
In the context of dimensional regularization (DREG)~\cite{Bollini:1972ui, 'tHooft:1972fi}, 
we work in conventional dimensional regularization (CDR) to make explicit the IR divergent 
structure, with $d=4-2\ep$ the number of space-time dimensions. 
The corresponding squared splitting amplitudes
in other schemes can be obtained by computing scalar-gluon contributions or 
by using transition rules if we are only interested in the ${\cal O}(\ep^0)$ terms. Transition rules for scattering amplitudes were previously discussed in the literature \cite{Kunszt:1993sd, Draggiotis:2009yb, Garzelli:2009is}, 
while we treated this topic for the double-collinear limit in Ref.~\cite{Sborlini:2013jba}.
We will not enter into this discussion in this paper.

The outline of the paper is the following. In Section~\ref{sec:multiple}
we describe the 
multiple-collinear kinematics and introduce some relevant properties and notation. 
After some brief comments about QCD with photons, we review the double-collinear 
splitting functions for processes with photons (both at the
amplitude and squared-amplitude level). 
Then, we present explicit results for NLO unpolarized splitting kernels with one and 
two photons in Section~\ref{sec:partons}. 
To be more specific, we compute $q \to q \gamma \gamma$, 
$q \to q g \gamma$ and $g \to q \bar{q} \gamma$ at NLO in QCD. 
In Section~\ref{sec:photons} we show results for triple 
splitting kernels involving one photon which splits into three partons: 
$\gamma \to q \bar{q} \gamma$ and $\gamma \to q \bar{q} g$. 
In this section, we also present a discussion about the functional 
structure of these objects, using the photon-started splitting processes 
to handle more compact expressions. 
Finally we present the conclusions and perspectives in Section~\ref{sec:conclusions}.

\section{Multiple-collinear limit and IR divergent behavior}
\label{sec:multiple}

To explore the multiple collinear limit, we consider an $n$-particle process 
in which a subset of $m$ particles become simultaneously collinear. 
We label the collinear particles with $i\in C=\{1,2, \ldots, m\}$. 
If $p_i^{\mu}$ is the four-momentum associated with particle $i$, 
and assuming $p_i^2=0$ (massless on-shell partons),
then the subenergies $s_{i j} = 2 \, p_i \cdot p_j$ and 
$s_{i,j} = \left(p_i +p_{i+1} + \ldots + p_{j} \right)^2 = p_{i,j}^2$, 
with $i$ and $j \in C$, are the only dimensional relevant quantities 
involved in the collinear limit.
We define a pair of light-like vectors ($\wp^2 = 0$, $n^2 = 0$), 
such that 
\beq
\wp^{\mu} = p_{1,m}^{\mu} - \frac{s_{1,m}}{2 \; n \cdot \wp} \, n^{\mu}
\label{ptilde}
\eeq
approaches the collinear direction in the multiparton collinear limit, 
and $n^\mu$ parametrizes how the collinear 
limit is approached, with $n\cdot\wp = n\cdot p_{1,m}$.
The longitudinal-momentum fractions $z_i$ of the collinear partons are
\beq
z_i = \frac{n \cdotp p_i}{n\cdot \wp}~, \qquad i \in C~,
\eeq
and fulfil the constraint $\sum_{i\in C} z_i = 1$.
We work in the physical-gauge where the factorization properties are
explicit, with 
\beq
d_{\mu \nu} (k,n) = - \eta_{\mu \nu} + \frac{k_{\mu} n_{\nu} + n_{\mu} k_{\nu}}{n\cdot k}
\label{bosonpol}
\eeq 
the physical polarization tensor of a gauge vector boson (gluon or photon) 
with momentum $k$. The auxiliary gauge vector $n$ in \Eq{bosonpol} is 
taken identical to the light-like vector $n$ introduced in \Eq{ptilde}. 

On the other hand, let's recall some universal properties of the collinear 
limit of scattering amplitudes. 
Strict collinear factorization is fulfilled to all orders in the time-like region 
(i.e. where $s_{ij}>0$ for all $i,j \in C$). For simplicity, we limit ourselves to this region 
for the rest of the paper and leave the computation of factorization breaking 
effects in the space-like region~\cite{Catani:2011st,Forshaw:2012bi} for a future publication. 
When particles $1$ to $m$ become collinear, the one-loop matrix element ${\cal M}^{(1)}$ 
factorizes according to
\beqn
\nn 
\ket{{\cal M}^{(1)}\left(p_1, \ldots, p_n\right)} &\simeq& 
\sp^{(1)}_{a_1 \dots a_m}(p_1 , \ldots, p_m; \wp) \, 
\ket{{\cal M}^{(0)}(\wp, p_{m+1}, \ldots, p_n)} 
\\ &+& \sp^{(0)}_{a_1 \dots a_m}(p_1 , \ldots, p_m; \wp) \,
\ket{{\cal M}^{(1)}(\wp, p_{m+1}, \ldots, p_n)}~,  
\eeqn
which is valid for the most singular terms in the collinear limit.
The factors $\sp^{(0)}$ and $\sp^{(1)}$, also called splitting matrices, encode the singular
behavior in the multiple collinear limit at tree-level and one-loop, respectively.
As discussed in Ref.~\cite{Catani:2011st}, they turn out to be process-independent 
in the time-like region as they depend on the momenta and quantum 
numbers (flavour, spin, and color) of the collinear partons only. 
Centering in $\sp^{(1)}$, it can be expressed as
\beqn
\sp^{(1)}_{a_1 \dots a_m} &=& \sp^{(1)\,{\rm div.}}_{a_1 \dots a_m} + \sp^{(1)\,{\rm fin.}}_{a_1 \dots a_m}~,
\label{DescomposicionSP}
\eeqn
where all the infrared divergences are contained in $\sp^{(1)\,{\rm div.}}$, 
which can be expanded as
\beq 
\sp^{(1)\,{\rm div.}}_{a_1 \dots a_m} (p_1,\dots,p_m; \wp) =
\IC^{(1)}_{a_1 \dots a_m}(p_1,\dots,p_m;\wp) \, \sp^{(0)}_{a_1 \dots a_m}(p_1, \dots, p_m; \wp)~,
\label{DescomposicionSP2}
\eeq
with
\beeq 
\IC^{(1)}_{a_1 \dots a_m}(p_1,\dots,p_m;\wp) &=& \CG \, \g^2 \, \left( \frac{-s_{1, m} -i0}{\mu^2}\right)^{-\ep} \, 
\Bigg\{ \frac{1}{\ep^2}
\sum_{i,j=1 (i \neq j)}^{\bar{m}} \;{\bom T}_i \cdot {\bom T}_j
\left( \frac{-s_{ij} -i0}{-s_{1, m} -i0}\right)^{-\ep} \nn \\
&+&
\frac{1}{\ep^2}
\sum_{i,j=1}^{\bar{m}} \;{\bom T}_i \cdot {\bom T}_j
\;\left( 2 - \left( z_i \right)^{-\ep} -\left( z_j \right)^{-\ep} \right) \nn \\
&-& \frac{1}{\ep} 
\left( \sum_{i=1}^{\bar{m}} \left( \gamma_i - \ep {\tilde \gamma}_i^{\RS} \right)
- \left( \gamma_a - \ep {\tilde \gamma}_a^{\RS} 
\right) - \frac{\tilde{m}-2}{2} \left( \beta_0 - \ep {\tilde \beta}_0^{\RS}
\right) \right) \Bigg\}~,
\label{sp1div}
\eeeq
where the color matrix of the collinear particle with momentum $p_i$ is denoted by ${\bom T}_i$, 
$\bar{m}$ counts the number of collinear final state QCD partons and $\tilde{m}$ refers to the 
total number of QCD partons in the splitting process. 
We also define
\beq
\CG = \frac{\Gamma\left(1+\ep\right)\, \Gamma\left(1-\ep\right)^2}{\left(4 \pi\right)^{2-\ep}\, \Gamma\left(1-2\ep\right)} \, ,
\eeq
as the usual one-loop $d$-dimensional volume factor. We can appreciate that 
$\tilde{m}=\bar{m}$ in collinear splittings 
which are started by non-QCD partons. We order the final state particles 
such that $\left\{1, \ldots, \bar{m}\right\}$ 
are the colored ones while the remaining ones 
are singlets under ${\rm SU}(N_{\rm C})$ transformations.
The flavour coefficients $\gamma_i$ and $\beta_0$ are $\gamma_q=\gamma_{\bar q}=3C_F/2$ 
and $\gamma_g=\beta_0/2=(11C_A-2N_f)/6$. Up to ${\cal O}(\ep^0)$, 
the regularization scheme (RS) dependence is controlled by the coefficients 
${\tilde \gamma}_i^{\RS}$ and ${\tilde \beta}_0^{\RS}$. Explicitly, we have
\beq
{\tilde \gamma}_i^{\cdr}={\tilde \beta}_0^{\cdr}=0 \; ,
\eeq
in conventional dimensional regularization (CDR), while
\beqn
{\tilde \gamma}_q^{\dimr}&=&{\tilde \gamma}_{\bar q}^{\dimr}=C_F/2 \; , 
\\ {\tilde \gamma}_g^{\dimr}&=&{\tilde \beta}_0^{\dimr}/2=C_A/6 \; ,
\eeqn
in dimensional reduction (DR). Moreover, we consider unrenormalized quantities, 
so $\g$ is the bare QCD coupling. In the curly bracket of \Eq{sp1div}, the contribution 
proportional to $\beta_0 - \ep {\tilde \beta}_0^{\RS}$ is of ultraviolet origin, 
and it will be removed by working at the level of renormalized matrix elements 
and splitting matrices.

The factor $\IC^{(1)}_{a_1 \dots a_m}$ in \Eq{sp1div} 
is a matrix in the color space. However, since color conservation implies 
that the color charge of the parent parton is given by
\beq
\sum_i {\bom T}_i \;\sp^{(0)}_{a_1 \dots a_m} = \sp^{(0)}_{a_1 \dots a_m} \;{\bom T}_a~,
\eeq
and since we consider processes with at most $\tilde m = 3$ QCD partons,
the color algebra can be carried out in closed form and $\IC^{(1)}_{a_1 \dots a_m}$
becomes proportional to the unit matrix, namely, it becomes a $c$-number
that we denote $I^{(1)}_{a_1 \dots a_m}$.

The square of the splitting matrix $\sp_{a_1 \ldots a_m}$, 
summed over final-state colors and spins
and averaged over colors and spins of the parent parton, 
defines the $m$-parton unpolarized splitting function 
$\la \hat{P}_{a_1\ldots a_m} \ra$, which is a generalization of the customary 
Altarelli-Parisi double collinear splitting function. 
Fixing the normalization of the tree-level splitting function by
\beq
\label{p0def}
\la \hat{P}_{a_1 \cdots a_m}^{(0)} \ra = 
\left( \frac{s_{1,m}}{2 \;\mu^{2\ep}} \right)^{m-1} \,
{\overline {| \sp_{a_1 \dots a_m}^{(0)} |^2}}~,
\eeq
then it is possible to use \Eq{DescomposicionSP} and \Eq{DescomposicionSP2} 
to present the NLO correction as
\beqn
\nn \la \hat{P}_{a_1 \cdots a_m}^{(1)} \ra 
&=& \left( \frac{s_{1,m}}{2 \;\mu^{2\ep}} \right)^{m-1} \; \left(
{\overline { \sp_{a_1 \dots a_m}^{(1)} \left(\sp_{a_1 \dots a_m}^{(0)}\right)^\dagger }} + {\rm h.c.} \right)
\\ &=& 2 \, {\rm Re}\left( I^{(1)}_{a_1 \cdots a_m}(p_1,\dots,p_m;\wp) \right) 
\la \hat{P}_{a_1 \cdots a_m}^{(0)} \ra + \left( \la \hat{P}^{(1)\,{\rm fin.}}_{a_1 \cdots a_m} \ra +  {\rm c.c.}\right)~,
\label{EquacionDescomposicion1}
\eeqn
with the aim of exposing explicitly the divergent structure of the splitting function. 
Then, working with the finite remainder, we classify each term according to its 
transcendental weight. So, the second step of the decomposition that we carried out 
is to express the finite contribution to the NLO unpolarized splitting kernel as
\beqn
\la \hat{P}^{(1)\,{\rm fin.}}_{a_1 \cdots a_m} \ra
 &=& c^{a_1 \cdots a_m} \left[C^{(0)} + \sum^{2}_{i=1}\sum_{j} C^{(i)}_{j} F^{(i)}_{j}(\left\{s_{rl},z_k\right\}) \, + {\cal O}(\ep) \right]  \, ,
\label{EquacionDescomposicion2}
\eeqn 
where $F^{(i)}_{j}$ denotes a basis of a function's space with transcendental 
weight $i$ and $C^{(i)}_{j}$ are the corresponding coefficients. 
Here $c^{a_1 \cdots a_m}$ is a normalization factor which depends on the process 
and includes all the couplings.

\section{Splitting functions with photons}
\label{sec:partons}

In this section we present explicit results for the NLO corrections to triple collinear splitting functions involving photons. We describe the corresponding model through the Lagrangian density given by
\beqn
{\cal L}_{\rm QCD+QED}&=& {\cal L}_{\rm QCD} - \mu^{\ep} \, g_e \, \sum_q  
e_q \, \delta_{ij} \, \bar{\Psi}^i_q\gamma^{\mu}\Psi^j_q \, 
A_{\mu} -\frac{1}{4} F^{\mu\nu}F_{\mu\nu}~,
\eeqn
where $\left\{i,j\right\}$ are color indices, $g_e$ is the electromagnetic coupling (i.e. the absolute value of electron charge), $e_q$ is the charge of quark's flavour $q$ ($e_{u,c,t}=2/3$ and $e_{d,s,b}=-1/3$) and $F_{\mu \nu}=\partial_{\mu}A_{\nu}-\partial_{\nu}A_{\mu}$ is the gauge-field strength tensor for the Abelian group ${\rm U}(1)_E$. The photon-quark interaction is proportional to the identity matrix $\IDC$ in the color space.

Besides that, the computation is performed in the time-like region. With this kinematic choice, we can ensure strict factorization properties (as claimed in Ref.~\cite{Catani:2011st}). Centering in the triple collinear limit and using the notation previously introduced, we have $m=3$ but $\bar{m}$ is not fixed. Since we are interested in QCD corrections, 
we consider first splitting processes where the parent parton is a QCD parton. 
For $\bar{m}=1$, we have the splitting $a \to a \gamma \gamma$ with $a$ a quark or gluon. 
Since $\sp^{(0)}_{g \gamma \gamma} = 0$, the one-loop correction $\sp^{(1)}_{g \gamma \gamma}$
is ultraviolet finite and the only allowed unpolarized splitting function
at NLO with $\bar{m}=1$ is $\la \hat{P}^{(1)}_{q \gamma \gamma} \ra$.
When $\bar{m}=2$, with a photon in the final state, there are more possibilities. 
The relevant splittings functions are $\la \hat{P}_{q g \gamma} \ra$, 
$\la \hat{P}_{q \bar{q} \gamma} \ra$, and 
$\la \hat{P}_{g g \gamma} \ra$ where $\sp^{(0)}_{g g \gamma}=0$.
Finally, splitting processes initiated by photons will be discussed in 
Section~\ref{sec:photons}.

As mentioned in the introduction, the computation is performed using CDR scheme in order to simplify the treatment of intermediate expressions. It is possible to obtain DR scheme results up to ${\cal O}(\ep^0)$ by replacing the flavour coefficients and ${\tilde \beta}_0^{\RS}$ in \Eq{sp1div}. Also, we could obtain the exact expressions in other schemes by computing the corresponding scalar-gluon contributions \cite{Sborlini:2013jba}, although in this work we are only interested in ${\cal O}(\ep^0)$ corrections.

We start showing results for the $q \to q \gamma \gamma$ splitting function at NLO. Then, we express $q \to q g \gamma$ in terms of the previous one, in order to simplify the results. Finally, we compute $g \to q \bar{q} \gamma$ using the techniques mentioned above.

\subsection{Review of double-collinear results}
Before starting with the triple-collinear splitting functions, let's show some results for the double-collinear limit of scattering amplitudes with photons. This topic has been analysed in Ref.~\cite{Sborlini:2013jba} where we also discussed some technical details related with DREG schemes. In the context of QCD+QED, we have only two double-collinear splitting processes: $q \to q \gamma$ and $\gamma \to q \bar{q}$. 

\begin{figure}[htb]
	\centering
	\includegraphics[width=0.75\textwidth]{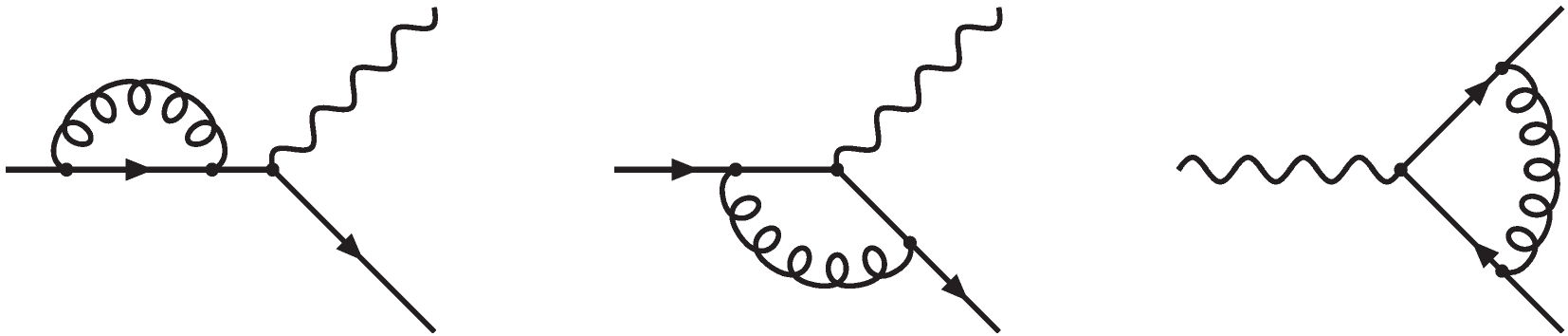}
	\caption{Feynman diagrams contributing to the splitting amplitudes of $q\to q \gamma$ and $\gamma\to q \bar{q}$. }
	\label{fig:Diagramas_DoblePHOTONES}
\end{figure}

The corresponding Feynman diagrams required to perform the computation are shown 
in Fig.~\ref{fig:Diagramas_DoblePHOTONES}. Using them, we obtained both splitting amplitudes and unpolarized splitting functions up to NLO accuracy. The results for the LO contributions are
\beqn
\sp^{(0)}_{q_1 \gamma_2} &=& \frac{g_e e_q \mu^{\ep}}{s_{12}} \IDC \,  \bar{u}(p_1)\slashed{\ep}{\left(p_2\right)}u(\wp) \ ,
\\ \sp^{(0)}_{\gamma \to q_1 	\bar{q}_2} &=&  \frac{g_e e_q \mu^{\ep}}{s_{12}} \IDC \, \bar{u}(p_1)\slashed{\ep}(\wp)v(p_2) \ ,
\eeqn
and
\beqn
\nn \sp^{(1)}_{q_1 \gamma_2} &=& \CG \g^2 C_F\, \left(\frac{-s_{12}-\imath 0}{\mu^{2}}\right)^{-\ep}\left[ \frac{2}{\ep^2}\left( 1-\, _2 F_1\left(1,-\ep ;1-\ep ;\frac{z_1-1}{z_1}\right) + \frac{\ep^2}{2(2\ep-1)}  \right)  \right.  
\sp^{(0)}_{q_1 \gamma_2}  \\ &-&  \frac{g_e e_q \mu^{\ep}}{s_{12} \, n\cdot \wp \, (2\ep-1)} \,  \IDC \, \bar{u}(p_1)\slashed{n}u(\wp) \, p_1\cdot \ep(p_2)  \bigg] \, ,
\\ \sp^{(1)}_{\gamma \to q_1 \bar{q}_2} &=& \CG \g^2 C_F \, \left(\frac{-s_{12}-\imath 0}{\mu^{2}}\right)^{-\ep}  \left[ -\frac{2}{\ep^2}-\frac{3}{\ep} + \frac{8}{2 \ep-1} \right] \sp^{(0)}_{\gamma \to q_1 \bar{q}_2} \, ,
\eeqn
for the QCD NLO corrections in the time-like region and using CDR scheme with 
$d=4-2\ep$. It is interesting to notice that $\gamma \to q \bar{q}$ only involves rational functions (aside from the global prefactor) and is extremely simple. In the last part of this article, after presenting the full NLO corrections to triple-splitting functions started by photons, we will discuss the origin of these compact expressions, but we can anticipate that it is related with the kind of integrals involved in the computation. In the double collinear limit, it is difficult to appreciate this fact because there is only one kinematical scale involved ($s_{12}$). In any case, since it is a three-particle process, computations can require triangle integrals at most. But only triangles with a LCG propagator can introduce hypergeometric functions which depend on $z_1$; otherwise the integral is completely independent of the momentum fraction. Moreover, the presence of momentum fractions inside non-rational functions is a consequence of having LCG propagators inside the loop integrals. Since this kind of functions is absent in $\sp^{(1)}_{\gamma \to q_1 \bar{q}_2}$, then all LCG propagators inside the loop should cancel. As a partial summary of the situation, there is a connection between gauge constraints and having on-shell QCD particles (charged under a \textit{non-Abelian} group) and just an off-shell parent photon (which is an \textit{Abelian} field).

For the sake of completeness, we show the corresponding splitting functions for the double-collinear limit with photons. At LO we have
\beqn
\la \hat{P}^{(0)}_{q_1 \gamma_2} \ra &=& g_e^2 e_q^2 \, \frac{1+z_1^2-\Delta (1-z_1)^2}{1-z_1} = g_e^2 e_q^2 \la {\cal{P}}_{q_1 \gamma_2}^{(0)} \ra \, ,
\\ \la \hat{P}^{(0)}_{\gamma \to q_1 \bar{q}_2} \ra &=&g_e^2 e_q^2\frac{ C_A (1-2(1-z_1)z_1-\Delta )}{1-\Delta} = \frac{g_e^2 e_q^2 C_A}{1-\Delta} \la {\cal{P}}_{\gamma \to q_1 \bar{q}_2}^{(0)} \ra\, ,
\eeqn
where $\Delta=\delta \ep$, with the parameter $\delta$ used to switch between FDH/HV ($\delta=0$) and CDR ($\delta=1$) schemes. Choosing CDR scheme, the NLO corrections can be written as
\beqn
\nn \la \hat{P}^{(1)}_{q_1 \gamma_2} \ra &=& \CG  \g^2 \, C_F \, 
{\left(\frac{-s_{12}-\imath 0}{\mu^2}\right)}^{-\ep}  
\left[ \frac{2}{\ep^2}\left( 1-\, _2 F_1\left(1,-\ep ;1-\ep ;\frac{z_1-1}{z_1}\right) + \frac{\ep^2}{2(2\ep-1)}  \right) \la \hat{P}^{(0)}_{q_1 \gamma_2} \ra \right.
\\ &+&   \left. g_e^2 e_q^2 \, \frac{ z_1(1+z_1)}{(1-z_1)(1-2\ep)} \right]\, + {\rm c.c.} \, ,
\\ \la \hat{P}^{(1)}_{\gamma \to q_1 \bar{q}_2} \ra &=& \CG  \g^2 \, C_F \, 
\left(\frac{-s_{12}-\imath 0}{\mu^2}\right)^{-\ep}  
\left[ -\frac{2}{\ep^2}-\frac{3}{\ep} + \frac{8}{2 \ep-1} \right]
\la \hat{P}^{(0)}_{\gamma \to q_1 \bar{q}_2} \ra \, + {\rm c.c.} \, ,
\eeqn
where we adapted the original notation of \cite{Sborlini:2013jba} in order to be compatible with the one used in this article.

\subsection{$q \to q \gamma \gamma$}
Let's consider the process $q \to q \gamma \gamma$. In terms of color structure, 
this is the simplest case because it is proportional to the identity matrix $\IDC$. 
The corresponding splitting amplitude at tree-level is given by
\beqn
\Spamplitude^{(0)(a_1;a)}_{q_1 \gamma_2 \gamma_3} &=& 
\frac{g_e^2 \, e_q^2 \, \mu^{2\ep} \, \IDC_{a_1 a}}{s_{123} \, s_{13} } \, \bar{u}(p_1) 
\left( \slashed{\ep}(p_3) \slashed{p}_3 + 2 \, p_1 \cdot \ep(p_3) \right) 
\slashed{\ep}(p_2) \, u(\wp) \, + (2\leftrightarrow 3 ) \, .
\label{SplittingLOq-qAA}
\eeqn

\begin{figure}[thb]
	\centering
	\includegraphics[width=\textwidth]{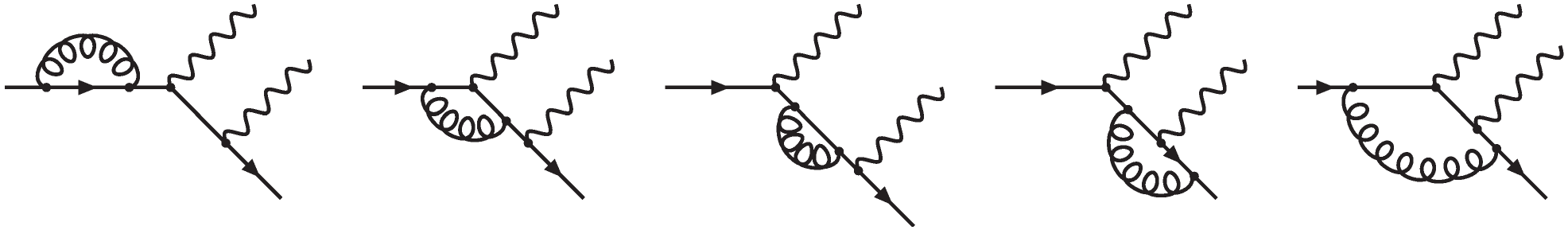}
	\caption{Representative Feynman diagrams contributing to the $q\to q \gamma \gamma$ splitting amplitude. 
        This splitting amplitude is symmetric under the exchange of the two final state photons.}
	\label{fig:Diagramas_q-q2A}
\end{figure}

In order to reduce the output size and to show explicitly that the splitting function is dimensionless, we use the notation
\beqn
x_{i} &=& \frac{-s_{j k}- \imath 0}{-s_{123}- \imath 0}~,
\eeqn
with $(i,j,k)$ a reordering of the indices set $\left\{1,2,3\right\}$ and the special case $x_0 \equiv 1$. Moreover, we define
\beqn
\Delta_{i',j'}^{i,j} &\equiv& x_{i} z_j + x_{i'} (z_{j'}-1) \, ,
\\ \bar{\Delta}_{i',j'}^{i,j} &\equiv& x_{i}z_j - x_{i'}(z_{j'}-1) \, ,
\eeqn
where the indices represent outgoing particles, and $z_0 \equiv 1$. Using this notation, it is possible to write the unpolarized splitting function at LO as
\beeq
\nn \la \hat{P}^{(0)}_{q_1 \gamma_2 \gamma_3} \ra &=& \frac{e^4_q g^4_e}{x_2}\left(\frac{\KerLO}{z_2}\left(1+ \frac{(1+x_2) z_1}{x_3}\right) +(\Delta -1) \left(\Delta  \left(x_1-\frac{z_1 (x_1+1)}{1-x_1}+\frac{z_1}{2 x_3}\right) \right.\right.
\\ &+& \left.\left. \vphantom{\frac{z_1 (2 (x_1+1) x_3-(1-x_1))}{2 (1-x_1) x_3}} x_3+z_1-z_2+2 \right) \right)  + \, (2 \leftrightarrow 3)\, .
\eeeq

Going to the NLO correction, we use the decomposition suggested in 
\Eq{EquacionDescomposicion1} and \Eq{EquacionDescomposicion2}.
The factor predicted by Catani's formula is 
\beqn
I^{(1)}_{q_1 \gamma_2 \gamma_3} (p_1, p_2, p_3; \wp) &=& 2 \CG \, \g^2 \, C_F \, 
\left(\frac{-s_{123}-\imath 0}{\mu^2}\right)^{-\ep } \frac{ 1-z_1^{-\ep }}{\ep ^2}~,
\eeqn 
and we found agreement with the divergent contribution in our results.

The rational coefficient is given by
\beqn
\nn C^{(0)} &=& \frac{\KerLO (z_1-1) (x_1 (x_2-1)+(x_3-1) z_2)}{x_1^2 x_2 (x_3-1) z_2} +\frac{(x_1-1) ({\Delta^{1,0}_{0,1}}-3)}{x_1 x_2 z_2}+\frac{3 x_1^2 \bar{\Delta}^{3,0}_{0,2}-z_2}{x_1^2 (1-x_2) x_2}
\\ \nn &+& \frac{(\Delta^{1,0}_{0,1})^2+2 x_1 z_2+2 (1-x_1)^2}{x_1^2 x_2}+\frac{\Delta^{1,2}_{0,1}-x_1 z_3}{x_1^2 (1-x_2)} +\frac{z_1 \left(x_1^2+2 (x_1+1) x_3\right)}{x_1^2 (x_3-1) x_3 z_2}
\\ &+&\frac{1}{x_1^2 z_2}\left(\frac{x_1^3}{1-x_2}+\frac{2 z_1}{x_2}\right) -\frac{3 (1-x_2) z_1}{x_1 x_2 x_3 z_2} +\frac{4 (z_1-1)}{x_1 z_2} \, +(2 \leftrightarrow 3) \, ,
\eeqn
with the global prefactor
\beqn
c^{q \gamma \gamma} &=& e_q^4 g_e^4 \g^2 C_{F} \, .
\eeqn

At transcendental weight 1, the basis is composed by 
\beqn
F^{(1)}_1 &=& \log\left(x_1\right) \,,
\\ F^{(1)}_2 &=& \log\left(x_3\right) \,. 
\eeqn
It is important to notice that these logarithms depend on ratios of kinematical variables, which ensures that this contribution can be integrated if we need to compute ${\rm N}^3$LO corrections. Due to symmetry considerations, we can represent the results as
\beqn
\la \hat{P}^{(1)\,{\rm fin.}}_{q_1 \gamma_2 \gamma_3} \ra
\left. \right|_{w=1} &=& c^{q \gamma \gamma} \left[\sum^2_{i=1}\, C^{(1)}_i F^{(1)}_i + \left(2\leftrightarrow 3 \right)   \right] \, ,
\eeqn
where
\beqn 
\nn C^{(1)}_1 &=& \frac{2 \Delta^{0,1}_{2,3} ((z_3-1) \Delta^{0,1}_{2,3}+x_3 (z_1+z_3 (2 z_2+z_3-2)+1))}{(1-x_1) x_3^2 z_2 z_3}+\frac{4 x_2 \bar{\Delta}^{3,0}_{0,3}}{x_3(1-x_1)}
\\ &+& \frac{2 z_2 z_3 (x_1+z_1)+(z_1+1) z_1}{(x_1-1) z_2 z_3} \, ,
\\ \nn C^{(1)}_2 &=& \frac{{\Delta^{1,0}_{0,1}} }{1-x_3} \left(\frac{2 ({\Delta^{1,0}_{0,1}}-z_1+1)}{x_1 x_2 (1-z_1) z_2}+\frac{{\Delta^{1,0}_{0,1}}+2 z_1}{x_1 x_3 z_3}+\frac{1}{x_1 z_2}-\frac{x_2 (1-z_1)-(1-x_3) (5 z_1-2)}{x_2 (1-x_3) (1-z_1) z_2}\right.
\\ \nn &-& \left. \frac{\Delta^{1,0}_{0,1}}{(1-z_1) z_2}\left(\frac{z_1 (3 x_1+4 x_2)+x_1-z_1+1}{x_1 x_2 (1-x_3)}+\frac{2}{x_2^2}\right) -4 \right)
\\ \nn &+& \frac{1}{1-x_3}\left(2 {\Delta^{2,0}_{0,2}}-\frac{3 x_1^2 {\Delta^{1,0}_{0,3}}+x_1 (z_2-z_3)+z_2+2 z_3}{x_1^2 (1-x_3)}+\frac{1}{z_2}\left(\frac{3 (1-z_1) z_1}{x_1}-\frac{2 {\Delta^{1,1}_{0,1}}}{x_1 (1-x_3)}\right) \right.
\\ \nn &+& \left. \frac{x_1 \left(x_1^2 (z_1-z_2+2)-x_1 (z_3+1)+4 z_1+z_2-2\right)-2 z_1-z_2+2}{x_1^2 x_2} + \frac{x_1}{(1-x_3) z_3} \right.
\\ \nn &+& \left. \frac{x_1 z_1^2 (2 x_1 z_1+x_2 (z_1-2))}{x_2^2 (z_1-1) z_3}-\frac{(x_1+3) z_1^2-4 z_1+1}{x_1 z_3}-2\right)
\\ &+& \frac{1-x_1}{1-x_3}\left(\frac{2 (x_1+z_1)-z_2-{\bar{\Delta}^{1,3}_{0,1}}}{x_1 x_3}-\frac{2 ({\Delta^{1,1}_{0,1}}-{\bar{\Delta}^{1,2}_{0,2}}+2 x_1)}{x_2^2}\right) \, .
\eeqn
The splitting is fully symmetric under the exchange of particles 2 and 3, then let's exploit this symmetry 
to minimize the basis. At weight 2:
\beqn
\la \hat{P}^{(1)\,{\rm fin.}}_{q_1 \gamma_2 \gamma_3} \ra
\left. \right|_{w=2} &=& c^{q \gamma \gamma} \left[ \sum^{7}_{i=1} C^{(2)}_i F_i^{(2)} \ + (2\leftrightarrow 3) \right] \, ,
\eeqn
where the corresponding basis is
\beqn
F^{(2)}_1 &=& \log^2 \left(z_1\right) \, ,
\\ F^{(2)}_2 &=& \frac{\pi^2}{6}-\li{1-x_1} \, , 
\\ F^{(2)}_3 &=& \log(x_1)\log(z_1) \, ,
\\ F^{(2)}_4 &=& \li{1-x_2} + \log(x_1)\log(x_2) \, ,
\\ F^{(2)}_5 &=&  {\cal R}\left(x_1,x_2\right) + 2\li{1-x_2} + 2\log(x_1)\log(x_2)\, ,
\\ \nn F^{(2)}_6 &=&  \li{\frac{\Delta^{2,0}_{0,2}}{z_2-1}}-\li{1-x_2}-\li{-\frac{z_3}{z_1}}-\li{\frac{z_2}{z_2-1}}
\\ &-& \log (x_2) \log \left(\frac{1-z_2}{z_1}\right) \, ,
\\ F^{(2)}_7 &=&  \li{z_2}+\log \left(\frac{1-z_2}{z_1}\right) \log \left(\frac{x_1 (1-z_2)}{x_2}\right)+\log (x_1) \log (z_1) \, ,
\eeqn
with $\Delta^{2,0}_{0,2} = x_2 + z_2 - 1$. The first three components of the basis are symmetric under the exchange of particles 2 and 3. On the other hand, in the previous list we introduced the function 
\beqn
{\cal R} \left(x_1,x_2\right) &=& \frac{\pi^2}{6}-\log (x_1) \log (x_2) - \li{1-x_1} - \li{1-x_2} \, ,
\eeqn
which is going to appear in all the remaining splitting functions, and whose origin will be explained in the last part of this article. Also, we can appreciate that ${\cal R}\left(x_i,x_j\right)=0$ when $x_k \to 0$ ($k \neq i,j$). Following with the presentation of the results, the coefficients associated with the previous basis are given by
\beqn
C^{(2)}_1 &=&  \frac{2 z_1 \left(z_2 z_3+(1-z_2)^2\right)}{x_2 x_3 z_2 z_3} \, ,
\\ \nn C^{(2)}_2 &=& \frac{4 (1-z_2) (\bar{\Delta}^{0,3}_{1,2})^2}{x_2^3 z_2 z_3}-\frac{16 x_1 (1-z_2)^2}{x_2^3 z_2}-\frac{2 (1-x_1)^2 \left(3 x_1 z_2 z_3+z_1 \left(z_1^2+1\right)\right)}{x_1 x_2 x_3 z_2 z_3}
\\ &+& \frac{4 z_1 (2 (1-z_2) z_3-\bar{\Delta}^{0,3}_{1,2})}{x_2^2 z_2 z_3}+\frac{4 \left(z_2 (x_1-2 z_1-z_3)+(1-z_3)^2\right)}{x_3 z_2}+\frac{4 (1-z_3)^2}{x_2 z_2} \, ,
\\ C^{(2)}_3 &=& \frac{4 \left(z_1 (1-z_2)+(1-z_3)^3\right)}{x_3 z_2 z_3}-\frac{4 x_2}{x_3} \, ,
\eeqn
for the explicitly symmetric contribution, and
\beqn
C^{(2)}_4 &=&  -\frac{2 (x_3 z_1-\Delta^{2,0}_{0,2}-x_1 z_3) \left(\left(z_1^2+1\right) \Delta^{2,2}_{3,2}-x_1 z_2 z_3 (2 x_2+x_3)\right)}{x_1 x_2 x_3 z_2 z_3 \Delta^{2,0}_{0,2}} \, ,
\\ \nn C^{(2)}_5 &=& -\frac{2 \KerLO (x_2 z_2 \bar{\Delta}^{1,0}_{0,1}+x_3 (z_1-1) (x_1 (z_2-1)+z_1-z_2+1))}{x_1 x_2 x_3 z_2 z_3}+\frac{2 \Delta^{1,0}_{0,1} (\Delta^{1,0}_{0,1}-x_3 z_1)}{x_3^3 (z_1-1) z_3}
\\ \nn &-& \frac{2 x_1^2 z_1^3-2 x_1 x_3 z_1^2+2 x_3^2 (z_1+1) z_1}{x_3^3 (1-z_1) z_2}+\frac{4 (1-x_1) (2 x_2+x_3)}{x_2 x_3}
\\ &-& \frac{2 (1-x_1) (x_1 (2 z_1+z_2+1)-z_2-1)}{x_3^3}+\frac{4 z_1}{x_3^2}+\frac{2}{x_3 z_3}-4\, ,
\\ C^{(2)}_6 &=& -\frac{2 \KerLO (1-z_1) ((z_1-1) \Delta^{3,0}_{0,3}+x_1 (z_1-z_2)-z_1 z_3)}{x_1 x_2 x_3 z_2 z_3}\, ,
\\ \nn C^{(2)}_7 &=& \frac{2\KerLO}{x_2 x_3 \Delta^{2,0}_{0,2}}\left(\frac{x_2 \Delta^{1,0}_{0,1}+x_2 x_3 (1-z_1)+2 x_3 z_1^2}{z_3}+\frac{x_1 x_2 z_1+x_2 x_3 (z_1-1)+2 x_3}{z_2}\right)
\\ \nn &-&\frac{2 \left(x_1 (z_2-1)+2 z_1^2+z_1 (3-2 z_2)+2 z_2^2-7 z_2+7\right)}{x_2 \Delta^{2,0}_{0,2}}+\frac{4 \Delta^{0,2}_{1,3}}{x_3 \Delta^{2,0}_{0,2}}
\\ &+& \frac{2 (z_1-2 z_2+3)}{\Delta^{2,0}_{0,2}} \, ,
\eeqn
for the others.


\subsection{$q \to q g \gamma$}
\begin{figure}[htb]
	\centering
	\includegraphics[width=0.8\textwidth]{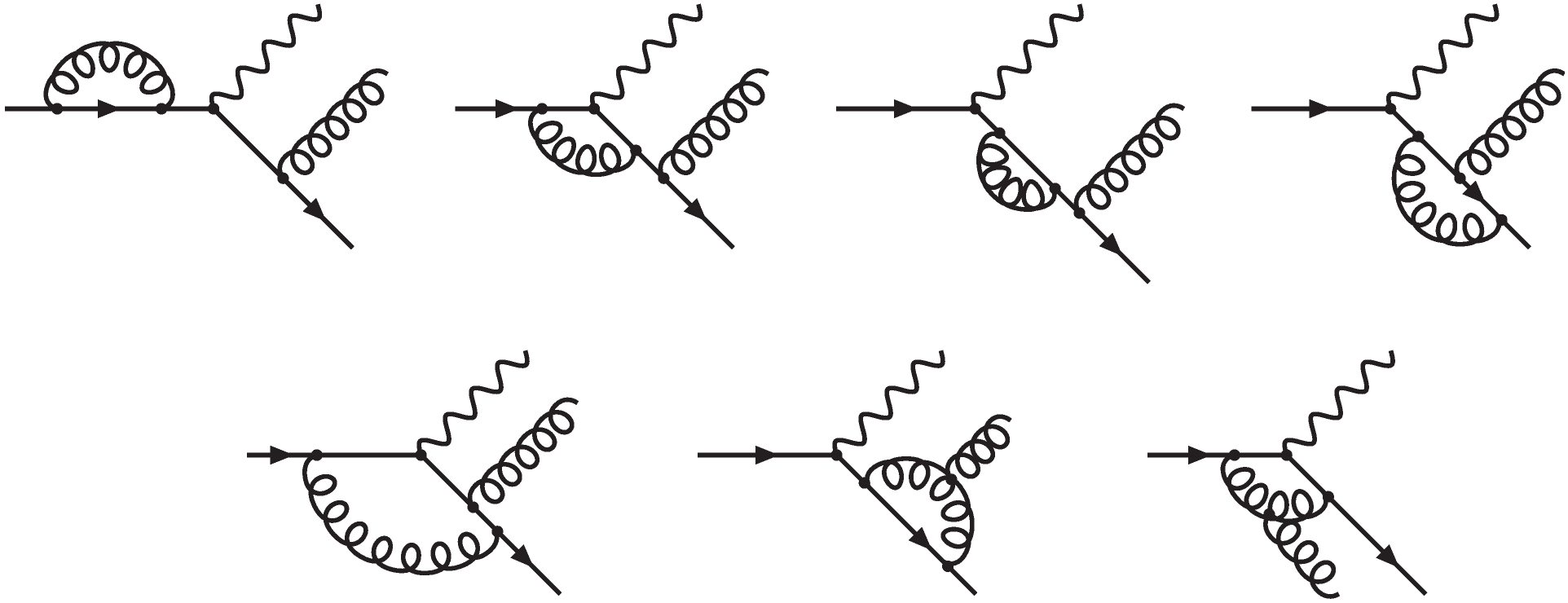}
	\caption{Representative Feynman diagrams contributing to the $q\to q g  \gamma$ splitting amplitude. 
The first 5 diagrams are associated with the \textit{Abelian} contribution already presented in $q\to q \gamma \gamma$. 
The last 2 diagrams lead to \textit{non-Abelian} (NA) contributions.}
	\label{fig:Diagramas_q-qgA}
\end{figure}

The natural following step is to replace one photon by a gluon. This will produce more complicated expressions, but it is possible to relate them with the $q \to q \gamma \gamma$ splitting. In order to do that, let's take a look at Fig. \ref{fig:Diagramas_q-qgA}. At LO, we have exactly the same kinematic structure, which implies that
\beqn
\nn \Spamplitude^{(0)(a_1,\alpha_2;a)}_{q_1 g_2 \gamma_3} &=& \frac{e_q g_e \mu^{2\ep} \SUNT^{\alpha_2}_{a_1 a}}{ {s_{123}}} \bar{u}(p_1) \left[
\left( \slashed{\ep}(p_3) \slashed{p}_3 + 2 \, p_1 \cdot \ep(p_3) \right) 
 \frac{\slashed{\ep}(p_2)}{s_{13}}  \, + (2\leftrightarrow 3 ) \, \right]\, u(\wp)
\\ &=& \SUNT^{\alpha_2}_{a_1 a} \, \frac{\g}{e_q g_e} \Spamplitude^{(0)(a_1;a)}_{q_1 \gamma_2 \gamma_3}\, ,
\label{SplittingLOq-qgA}
\eeqn
and the corresponding unpolarized LO splitting function is simply
\beqn
\la \hat{P}_{q_1 g_2 \gamma_3}^{(0)} \ra&=& C_F \, \frac{\g^2}{e^2_q g^2_e}  
\la \hat{P}_{q_1 \gamma_2 \gamma_3}^{(0)} \ra \, ,
\eeqn
in terms of the one associated to $q \to q \gamma \gamma$.

To obtain the NLO corrections, we use the decomposition suggested in 
\Eq{EquacionDescomposicion1} and \Eq{EquacionDescomposicion2}. 
The factor predicted by Catani's formula is
\beqn
I^{(1)}_{q_1 g_2 \gamma_3} (p_1, p_2, p_3; \wp) &=& \frac{\CG \g^2}{\ep^2} \left(\frac{-s_{123}-\imath 0}{\mu^2}\right)^{-\ep } \left[2 C_F - C_A \left(x_3^{-\ep} +z_2^{-\ep}\right) - D_A z_1^{-\ep}\right]
\\ \nn &=& \frac{D_A}{2 C_F} I^{(1)}_{q_1 \gamma_2 \gamma_3} \left(p_1,p_2,p_3;\wp \right) +  
C_A \, I^{(1,C_A)}_{q_1 g_2 \gamma_3} (p_1, p_2, p_3; \wp)~,
\eeqn 
where 
\beq
I^{(1,C_A)}_{q_1 g_2 \gamma_3} (p_1, p_2, p_3; \wp) =
\frac{\CG \g^2}{\ep^2} \, \left(\frac{-s_{123}-\imath 0}{\mu^2}\right)^{-\ep } 
\left(1-x_3^{-\ep} -z_2^{-\ep}\right)~,
\eeq
with $D_A = 2 C_F-C_A$ and we found agreement with the divergent contribution 
in our results.

Before writing explicit results, let's take a look to the corresponding NLO Feynman diagrams contributing to the splitting amplitude. If we just consider the topology, we can appreciate that many diagrams are already present in Fig. \ref{fig:Diagramas_q-q2A}. Since a photon has been replaced by a gluon, there is an extra color matrix. But in these diagrams, the kinematic structure is unchanged and only some color factors need to be modified. At amplitude level, in $q \to q \gamma \gamma$ there was a global color factor $\left(\SUNT^\beta \SUNT^\beta\right)_{a_1 a} = C_F \, \IDC_{a_1 a}$
related with a gluon of color $\beta$ inside the loop. However, in $q \to q g \gamma$ we must insert the color matrix $\SUNT^\alpha$ associated with the external gluon, which leads to two possibilities: 
\begin{itemize}
	\item $\left(\SUNT^\beta \SUNT^\beta \SUNT^\alpha\right)_{a_1 a} = \frac{D_A+C_A}{2} \, \SUNT^\alpha_{a_1 a}$ if the external gluon is attached to a fermion line inside the loop, or;
	\item $\left(\SUNT^\beta \SUNT^\alpha \SUNT^\beta\right)_{a_1 a} = \frac{D_A}{2} \, \SUNT^\alpha_{a_1 a}$ when the external gluon interacts with an external fermion (i.e. a fermionic line which is not inside a loop).
\end{itemize}
It is important to notice that both configurations include a contribution proportional to $C_F$. 
On the other hand, there are some diagrams which include a triple-gluon vertex in 
Fig.~\ref{fig:Diagramas_q-qgA} whose topology is different from the ones associated 
with the $q \to q \gamma \gamma$ process.

In summary, we conclude that it is possible to express the $q \to q g \gamma$ splitting function in terms of $q \to q \gamma \gamma$. To make this relation explicit at NLO, we write
\beqn
\la \hat{P}_{q_1 g_2 \gamma_3}^{(1)} \ra&=& D_A \la \hat{P}^{(1,D_A)}_{q_1 g_2 \gamma_3} \ra + C_A \la \hat{P}^{(1,C_A)}_{q_1 g_2 \gamma_3} \ra \; + \; \rm c.c. \, ,
\eeqn
with 
\beqn
\la \hat{P}^{(1,D_A)}_{q_1 g_2 \gamma_3} \ra &=& \frac{1}{2 C_F} \left[
{\rm Re} \left( I^{(1)}_{q_1 \gamma_2 \gamma_3} (p_1,p_2,p_3;\wp) \right) 
\la \hat{P}_{q_1 g_2 \gamma_3}^{(0)} \ra + 
\la \hat{P}^{(1)\,{\rm fin.}}_{q_1 \gamma_2 \gamma_3} \ra \right]~,
\\ \la \hat{P}^{(1,C_A)}_{q_1 g_2 \gamma_3} \ra &=& 
{\rm Re} \left( I^{(1,C_A)}_{q_1 g_2 \gamma_3} (p_1, p_2, p_3; \wp) \right) \,
\la \hat{P}^{(0)}_{q_1 g_2 \gamma_3} \ra 
+ \la \hat{P}^{(1,C_A)\,{\rm fin.}}_{q_1 g_2 \gamma_3} \ra~,
\eeqn
where the global prefactor is given by
\beqn
c^{q g \gamma} &=& e_q^2 g_e^2 \g^4 C_{F} \, .
\eeqn
Centering in the finite NLO contribution, we can write the corrections proportional to $C_A$ using \Eq{EquacionDescomposicion2} as
\beqn
\la \hat{P}^{(1,C_A)\,{\rm fin.}}_{q_1 g_2 \gamma_3} \ra &=& c^{q g \gamma} \left[ C^{(0,C_A)} 
+ \sum^{2}_{i=1}\sum_{j} C^{(i,C_A)}_{j} F^{(i,C_A)}_{j}(\left\{x_{l},z_k\right\}) \, + {\cal O}(\ep) \right]  \, .
\eeqn

After these considerations, let's show the explicit results. In first place, the rational coefficient is given by
\beqn
\nn C^{(0,C_A)} &=&  \frac{x_1 (z_2 (\Delta^{3,0}_{0,3}+4 (1-x_2 z_3))+3 x_3 z_3)+x_2 z_3 (3 z_2-4 x_3)}{2 x_1 (x_2-1) x_2 z_3}+\frac{(2-x_1) x_1^2+x_2 (x_3-1) z_2}{2 x_1 (x_2-1) x_2 x_3 z_2^{-1} z_3}
\\ &-& \frac{(1-x_3) (4 (1-x_2)-x_1 (3 x_1+x_3))-x_1 (x_1+1) x_3}{2 (x_2-1) x_2 (x_3-1) x_3 z_2} - (2 \leftrightarrow 3) \, ,
\eeqn
which is totally antisymmetric when interchanging particles 2 and 3. Since the rational contribution for the $q \to q \gamma \gamma$ kernel is totally symmetric, it is clear that $C^{(0,C_A)}$ is related with the NA diagrams associated with the non-Abelian triple-vertex.

At weight 1, the basis is composed by 
\beqn
F^{(1,C_A)} &=& \log\left(x_{3}\right)  \, ,
\eeqn
which only depend on ratios of kinematical variables. Due to symmetry considerations, we can rewrite this contribution as
\beqn
\la \hat{P}^{(1,C_A)\,{\rm fin.}}_{q_1 g_2 \gamma_3} \ra \left. \right|_{w=1} &=& c^{q g \gamma}  \left[ \left( C^{(1,C_A)}_{\rm sym} F^{(1,C_A)} +  (2\leftrightarrow 3 ) \right) \, + \left( C^{(1,C_A)}_{\rm asym} F^{(1,C_A)} -  (2\leftrightarrow 3 ) \right) \right]\, ,
\eeqn
where
\beqn
\nn C^{(1,C_A)}_{\rm sym} &=& \frac{3 (x_1 (1-{\Delta^{1,0}_{0,1}}-{\bar{\Delta}^{1,1}_{1,2}}-z_2)+z_1)}{2 x_1 x_2 (1-x_3)}+\frac{3 \left(z_3 {\Delta^{1,1}_{3,1}} \left({\Delta^{1,0}_{0,1}}-z_1^2+1\right)+x_1^2 z_1^3 z_2\right)}{2 x_1 x_2 (x_3-1) (z_1-1) z_2 z_3}
\\ &+& \frac{3 z_1 ({\Delta^{1,0}_{0,1}}-x_3 {\Delta^{1,1}_{0,1}})}{2 x_1 (1-x_3) x_3 z_3}+\frac{3 (2 {\Delta^{2,0}_{0,2}}-z_1-1)}{2 (1-x_3)}+\frac{3 (1-x_1) z_1}{2 x_1 (1-x_3) x_3} \, ,
\\ \nn C^{(1,C_A)}_{\rm asym} &=& \frac{\KerLO (z_1-1) }{x_1^2 x_3 z_3}\left(\frac{x_2}{1-x_3}-x_2-1\right) + \frac{(2 x_1-3) ({\Delta^{0,2}_{1,3}}+2 {\Delta^{1,0}_{0,1}})}{2 x_1^2 x_2 (1-x_3)}
\\ \nn &-& \frac{x_1 (3 x_1 {\Delta^{1,0}_{0,3}}+z_2-z_3)+z_2+2 z_3}{2 x_1^2 (1-x_3)^2}+\frac{(1-z_1) (2 ({\bar{\Delta}^{1,0}_{3,1}}-z_1+1)-5 x_1)}{2 x_1^2 (1-x_3) z_2}
\\ \nn &+& \frac{(x_1-1) \left(x_1^2 (z_3-2)-x_1 (2 z_1+z_2+1)-2 z_1+2 z_2+2\right)}{2 x_1^2 (1-x_3) x_3}
\\ \nn &+& \frac{x_1 (z_1-1) (2 x_3 z_1+x_3-3)-4 z_1 (x_2 x_3+x_3-1)}{2 x_1^2 (x_3-1) x_3 z_3}
\\ \nn &-& \frac{x_2 {\Delta^{1,1}_{0,1}} {\bar{\Delta}^{1,0}_{0,1}}+(2 x_1-3) (1-x_3) (1-z_1) {\Delta^{1,0}_{0,1}}}{2 x_1^2 x_2 (1-x_3)^2 z_2}
\\  &+& \frac{x_1+(x_3-1) (x_3 (z_1-1) z_3-z_1)}{2 (x_3-1)^2 x_3 z_3} \, .
\eeqn

Finally, if we go to transcendental weight 2, it is possible to perform the expansion
\beqn
\la \hat{P}^{(1,C_A)\,{\rm fin.}}_{q_1 g_2 \gamma_3} \ra
\left. \right|_{w=2} &=& c^{q g \gamma}  \, \sum^{7}_{i = 1} C^{(2,C_A)}_i F^{(2,C_A)}_i ~, 
\eeqn
using the set of functions
\beqn
F^{(2,C_A)}_1 &=& \li{z_3}+\log (1-z_3) \log \left(\frac{x_2 (1-z_3)}{z_2}\right)+\log (x_3) \log \left(\frac{z_2}{1-z_3}\right) \, ,
\\ F^{(2,C_A)}_2 &=& \frac{\pi ^2}{6}-\li{1-x_2} \, ,
\\ F^{(2,C_A)}_3 &=&  \li{1-x_3}-\li{z_3}+\log \left(\frac{1-z_3}{x_3}\right) \log \left(\frac{z_2}{x_2 (1-z_3)}\right) \, ,
\\ F^{(2,C_A)}_4 &=&  \frac{\pi ^2}{6}-\li{1-x_2}-\li{1-z_2}+\log (x_2) \log (z_2) \, ,
\\ \nn F^{(2,C_A)}_5 &=&  2\left[ \li{1-x_3}+ \li{-\frac{z_1}{z_2}}- \li{-\frac{\Delta^{3,0}_{0,3}}{1-z_3}}\right] -\li{z_3}
\\ &+& \log (x_2) \log (1-z_3)+\log \left(\frac{x_3}{z_2}\right) \log \left(\frac{1-z_3}{z_2}\right) \, ,
\\ F^{(2,C_A)}_6 &=&  \frac{\pi ^2}{6}-\li{1-x_2}-\li{1-z_2} \, ,
\\ F^{(2,C_A)}_7 &=&  {\cal R}\left(x_2,x_3\right)\, ,
\eeqn
where the associated coefficients are given by
\beqn
\nn C^{(2,C_A)}_1 &=& \frac{\KerLO }{1-x_1}\left(\frac{(x_1-3) z_1}{x_3 z_2}-\frac{1-x_1+2 z_1}{x_2 z_2}-\frac{(1-x_1) z_1+2}{x_2 z_3}+\frac{x_1-3}{x_3 z_3}\right)
\\ &+& \frac{4}{x_2}+\frac{2 (4-z_2-2 z_3)}{x_3} \, ,
\\ \nn C^{(2,C_A)}_2 &=&  \frac{\KerLO}{1-x_1}\left(\frac{3 x_1+z_1-3}{x_2 z_2}+\frac{(3 x_1-4) z_1+2}{x_2 z_3}+\frac{(3 x_1-2) z_1}{x_3 z_2}+\frac{3 x_1-z_1-1}{x_3 z_3}\right)
\\ &+& \frac{2 (2 (1-x_1+z_1)+z_2)}{x_3}-\frac{4 (x_1-z_1+z_2-2)}{x_2}-8 \, ,
\\ \nn C^{(2,C_A)}_3 &=& \frac{\KerLO }{(1-x_1) {\Delta^{3,0}_{0,3}}}\left(\frac{z_1 {\Delta^{1,0}_{0,1}}}{x_2 z_2}+\frac{{\bar{\Delta}^{1,0}_{1,1}}}{x_2 z_3}+\frac{{\bar{\Delta}^{1,0}_{0,1}}}{x_3 z_3}-\frac{\left(1-x_1^2\right) (1-z_1)}{x_2 x_3}+\frac{x_1 z_1^2}{x_3 z_2}\right)
\\ &-& \frac{2 (x_1 (z_2+1)-z_3)}{x_2 {\Delta^{3,0}_{0,3}}}-\frac{x_1 (1-z_3)+z_2}{x_3 {\Delta^{3,0}_{0,3}}}-\frac{1-z_1}{{\Delta^{3,0}_{0,3}}} \, ,
\\ \nn C^{(2,C_A)}_4 &=& \frac{\KerLO }{z_2 z_3 {\Delta^{2,0}_{0,2}}}\left((1-z_1) (z_3-z_2)-\frac{z_2 {\Delta^{1,0}_{0,1}}+x_1 z_1 z_3}{x_3}\right)+\frac{x_1 (z_2-1)+z_3}{x_2 {\Delta^{2,0}_{0,2}}}
\\ &-& \frac{2 \left((1-x_1) (2-x_1-z_3)-z_1+(1-z_3)^2\right)}{x_3 {\Delta^{2,0}_{0,2}}}+\frac{2 (2 x_2+3 z_2)-3 (1-z_1)}{{\Delta^{2,0}_{0,2}}}  \, ,
\\ C^{(2,C_A)}_5 &=& \frac{z_1 \left(z_1^2+1\right)}{2 x_2 x_3 z_2 z_3} + (2 \leftrightarrow 3) \, ,
\eeqn
\beqn
\nn C^{(2,C_A)}_6 &=& -\frac{2 \KerLO }{x_2 {\Delta^{2,0}_{0,2}}}\left(\frac{z_1^2}{z_3}+\frac{1}{z_2}\right)+\frac{2 \left((1-x_1)^2+(1-z_3)^2\right)}{x_3 {\Delta^{2,0}_{0,2}}}
\\ &-& \frac{4 {\bar{\Delta}^{2,0}_{0,3}}}{{\Delta^{2,0}_{0,2}}}+\frac{2 \left(z_1 (4-3 z_2)-z_2+z_3^2+2\right)}{x_2 {\Delta^{2,0}_{0,2}}} \, ,
\\ \nn C^{(2,C_A)}_7 &=&  \frac{\KerLO}{x_1^3 x_3}\left(\frac{x_1^3 (x_2 z_1+x_3)+x_3^3 (1-z_1)}{x_2 z_2}+\frac{{\Delta^{1,0}_{0,1}}+x_1^3 (z_1-x_1)+2 x_1}{z_3(1-x_1)}\right)
\\ \nn &-&  \frac{\KerLO}{z_3} \left(\frac{ {\Delta^{1,1}_{0,1}}+{\Delta^{1,1}_{2,1}}+x_1}{x_1^3} - \frac{{\Delta^{2,1}_{0,1}}+{\Delta^{3,1}_{0,1}}}{(1-x_1)x_2} \right) +\frac{2z_1\left( {\Delta^{1,1}_{0,1}}+{\Delta^{1,1}_{2,1}}+x_1\right)}{x_1^3 z_3 (1-z_1)}
\\ \nn &-&\frac{(1-x_1)^2 {\bar{\Delta}^{0,3}_{0,1}}+2 x_1^3 (x_3+z_1-z_2)}{x_1^3 x_2}+\frac{(1-x_1) (z_2-{\Delta^{1,0}_{0,1}}-{\bar{\Delta}^{1,1}_{1,3}})+2 x_1^3 (1-x_2)}{x_1^3 x_3}
\\ &+& \frac{x_1^2-2 x_3^2 z_1}{x_1^3 x_2 z_2}-\frac{x_1^3 (z_1+1)-4 x_1^2+6 x_1 z_1(z_1-1)^{-1}+2 z_1}{(x_1-1) x_1^3 x_3 z_3} \, .
\eeqn
Note that these coefficients mix both symmetric and antisymmetric contributions, as happened for the component of transcendental weight 1.


\subsection{$g \to q \bar{q} \gamma$}
Finally, we arrive to the last available configuration in the triple collinear limit with photons. Starting with the leading order, the splitting amplitude associated with the process $g \to q \bar{q} \gamma$ reads 
\beqn
\Spamplitude^{(0)(a_1,a_2;\alpha)}_{q_1 \bar{q}_2 \gamma_3} &=& \frac{e_q g_e \g \mu^{2\ep} \SUNT^{\alpha}_{a_1 a_2}}{ {s_{123}}} \, \bar{u}(p_1) 
\left( \frac{\slashed{\ep}(p_3) \slashed{p}_{13} \slashed{\ep}(\wp)}{s_{13}}-\frac{\slashed{\ep}(\wp) \slashed{p}_{23} \slashed{\ep}(p_3)}{s_{23}} \right) 
  \, v(p_2) \, \,
\, ,
\label{SplittingLOg-qqbA}
\eeqn
while the unpolarized splitting function is given by
\beqn
\nn \la \hat{P}_{q_1 \bar{q}_2 \gamma_3}^{(0)} \ra&=& \frac{e_q^2 g_e^2 \g^2}{2 x_1 x_2} \left[\left(\Delta^{1,0}_{0,1}\right)^2+z_1^2-\Delta  \left(\frac{2 z_2 (\Delta^{1,0}_{0,1}+1)+\Delta^{3,0}_{0,3}}{1-\Delta } + \frac{(1-x_3)^2}{2} \right) \right]
\\ &+& \, \, (1\leftrightarrow 2) \, ,
\eeqn
with $\Delta=\delta \ep$, as defined for $q \to q \gamma \gamma$ splitting. We can appreciate that this expression is totally symmetric when interchanging particles 1 and 2. At amplitude level, there is an additional minus sign coming from the change $e_q \to e_{\bar{q}}=- e_q$. On the other hand, it is useful to define
\beqn
\KerLOA &=& \frac{(\Delta^{1,0}_{0,1})^2 + z_1^2}{2 x_1 x_2} \, + (1\leftrightarrow 2) = \left. \frac{\la \hat{P}_{q_1 \bar{q}_2 \gamma_3}^{(0)} \ra}{e_q^2 g_e^2 \g^2} \right|_{\ep=0}\, ,
\eeqn
because this expression will allow to simplify NLO results.

\begin{figure}[htb]
	\centering
	\includegraphics[width=0.8\textwidth]{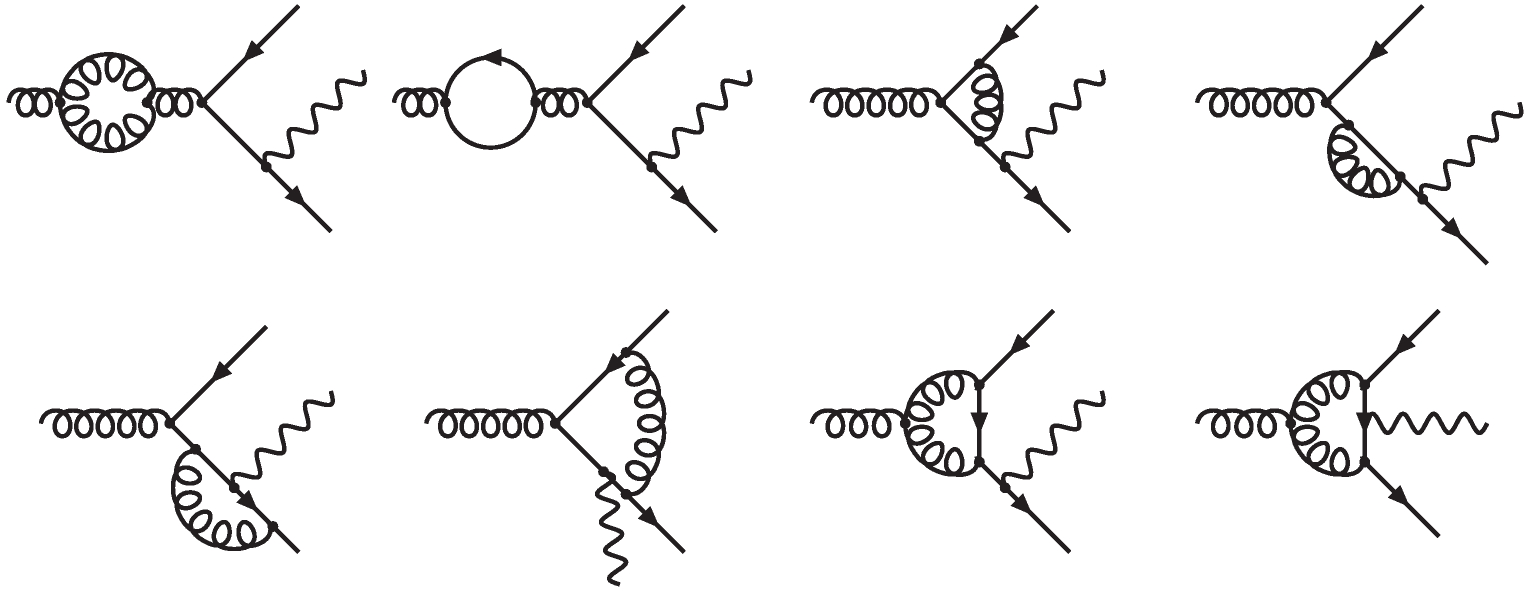}
	\caption{Representative diagrams contributing to the $g\to q \bar{q}  \gamma$ splitting amplitude.}
	\label{fig:Diagramas_g-qqbA}
\end{figure}

Before presenting the results, let's compare this process with $q \to q g \gamma$. The particles involved are the same, but we exchange the parent quark with a final state gluon. The corresponding Feynman diagrams are shown in Fig. \ref{fig:Diagramas_g-qqbA}. Excluding self-energy corrections to the parent parton, the other diagrams are in one-to-one correspondence in both processes. Moreover, both sets of diagrams are related by the exchange $P \leftrightarrow 2$. But the parent parton is an off-shell particle, and there are differences if process starts with a gluon or a quark. So, we do not expect to have any crossing transformation relating $\la \hat{P}_{q_1 \bar{q}_2 \gamma_3} \ra$ and $\la \hat{P}_{q_1 g_2 \gamma_3} \ra$ beyond tree-level. We will show a conclusive counterexample in the next section.

The IR divergent structure of this splitting is given by
\beqn
\nn I^{(1)}_{q_1 \bar{q}_2 \gamma_3} (p_1, p_2, p_3; \wp) &=& \frac{\CG \g^2}{\ep^2} \left(\frac{-s_{123}-\imath 0}{\mu^2}\right)^{-\ep } \left[ \vphantom{\left(2\gamma_q -\gamma_g - \frac{\beta_0}{2}\right)} C_A \left(2-z_1^{-\ep}-z_2^{-\ep}+x_3^{-\ep}\right)
\right. \\ &-& \left. 2 C_F \, x_3^{-\ep} - \ep \left(2\gamma_q -\gamma_g - \frac{\beta_0}{2}\right)\right] \, ,
\eeqn
according to Catani's formula and it agrees with our results.

To express the results in a compact way, we decompose it according to the color structure. So, the finite contribution to the unpolarized splitting function is expanded as
\beqn
\la \hat{P}_{q_1 \bar{q}_2 \gamma_3}^{(1)\,{\rm fin.}} \ra&=& c^{q \bar{q} \gamma}  \left[N_f \la \hat{P}^{(1,N_f)\,{\rm fin.}}_{q_1 \bar{q}_2 \gamma_3} \ra + D_A \la \hat{P}^{(1,D_A)\,{\rm fin.}}_{q_1 \bar{q}_2 \gamma_3} \ra + C_A \la \hat{P}^{(1,C_A)\,{\rm fin.}}_{q_1 \bar{q}_2 \gamma_3} \ra  \, + \, (1 \leftrightarrow 2) \right]\; \, ,
\eeqn
with the global prefactor
\beqn
c^{q \bar{q} \gamma} &=& \frac{e_q^2 g_e^2 \g^4}{4(1-\ep)} \, ,
\eeqn
and
\beqn
\la \hat{P}^{(1,N_f)\,{\rm fin.}}_{q_1 \bar{q}_2 \gamma_3} \ra &=& C^{(0,N_f)} \, ,
\\ \la \hat{P}^{(1,D_A)\,{\rm fin.}}_{q_1 \bar{q}_2 \gamma_3} \ra &=& C^{(0,D_A)} + \sum^{2}_{i=1} C_i^{(1,D_A)} F_i^{(1)} + C_1^{(2,D_A)} F_1^{(2,D_A)}  \, ,
\\ \la \hat{P}^{(1,C_A)\,{\rm fin.}}_{q_1 \bar{q}_2 \gamma_3} \ra &=& C^{(0,C_A)} + C_1^{(1,C_A)} F_1^{(1)} + \sum_{j=1}^{4}\,C_j^{(2,C_A)} F_j^{(2,C_A)}  \, ,
\eeqn
where we classify each contribution according to its transcendental weight. We can appreciate that the part proportional to $N_f$ is purely rational. Since it is originated from diagrams which contain fermionic loops, only \textit{standard} bubbles (i.e. without the extra LCG propagator) can contribute to $N_f$ terms. But these bubbles have single $\ep$ poles which are completely absorbed into $\IC^{(1)}_{q_1 \bar{q}_2 \gamma_3} (p_1, p_2, p_3; \wp)$, including also the corresponding logarithms. So, only the rational terms proportional to $N_f$ survive after the subtraction procedure.

Now let's show the explicit results. Starting with the rational terms, we have
\beqn
C^{(0,N_f)} &=& -\frac{20}{9} \KerLOA \, ,
\\ \nn C^{(0,C_A)} &=&  \frac{71}{9} \KerLOA + \frac{z_1 }{x_3}\left(\frac{x_3(1-2 z_3)+z_1}{x_1-1}+\frac{(x_3+1) (x_3-z_1)}{x_2 (x_3-1)}\right)
\\ &+& \frac{x_3 \left(z_1^2-z_1+1\right) (x_3-2 x_1+1)}{(1-x_1) x_1 (1-x_3)} \, ,
\\ \nn C^{(0,D_A)} &=&  (x_1 x_2-z_1 z_2-{\Delta^{1,0}_{0,1}}{\Delta^{2,0}_{0,2}})\left(\frac{2-2 x_1 (x_2+1)}{x_1 x_2 (1-x_3)}-\frac{1}{1-x_1}\right) 
\\ \nn &-& \frac{2 z_1 {\Delta^{1,0}_{0,1}} }{x_1 x_2}\left(\frac{(x_2+1) (1-x_2)^2}{(1-x_1) (1-x_3)}+\frac{x_3-x_1 x_2}{1-x_3}+\frac{(3-x_2) x_2-x_1 (x_2+1)}{2 (1-x_1)}+8\right) 
\\ &-& \frac{8 (1-x_1)^2}{x_1 x_2}-\frac{1-x_1}{x_1} \, ,
\eeqn
where we can appreciate that the expressions are symmetric under the transformation $1 \leftrightarrow 2$.

The basis for the contribution of transcendental weight 1 is given by
\beqn
F^{(1)}_{1} &=& \log\left(x_1\right) \, ,
\\ F^{(1)}_{2} &=& \log\left(x_3\right)  \, ,
\eeqn
and the associated coefficients are
\beqn
\nn C^{(1,D_A)}_{1} &=& \frac{x_1 x_2-z_1 z_2-{\Delta^{1,0}_{0,1}} {\Delta^{2,0}_{0,2}}}{(1-x_1) x_1}\left(\frac{x_2+2 x_3}{x_2^2}-\frac{x_1 x_2+2 x_3}{(1-x_1) x_2}-\frac{1+x_1}{1-x_1}\right)-\frac{z_2 (2 x_3-x_2) {\Delta^{2,0}_{0,2}}}{x_1 x_2^2}
\\ &-& \frac{(1-x_2)^2 z_1 (2 (1-x_1)+x_2) {\Delta^{1,0}_{0,1}}}{(1-x_1)^2 x_1 x_2^2}+\frac{(1-x_2) (x_2-2 x_3)}{(1-x_1) x_2} \, ,
\eeqn
\beqn
\nn C^{(1,D_A)}_{2} &=& \frac{2 \left(2 x_1 (z_2-1-{\Delta^{2,0}_{0,2}} (x_1 x_2+1))-{(\Delta^{0,3}_{1,2})}^2-2 x_2 (z_1+2 z_2-3) (x_1 x_2+2 z_3)-x_2 z_3\right)}{x_2^2 (1-x_3)^2}
\\ \nn &-&\frac{2 \left(x_1^2 (2 z_1 (z_2-3)+(4 z_2-13) z_2+7)+2 z_3^2\right)}{x_1 x_2 (1-x_3)^2}
\\ &-&\frac{2 (2 x_1 x_2+(z_1-15) z_1+7)}{(1-x_3)^2}  \, ,
\eeqn
for the Abelian part, and
\beqn
\nn C^{(1,C_A)}_{1} &=& \frac{z_2 (2 (x_2-2) z_1-9 x_2+10)+(x_2-1) (2 x_2 (z_1-2)-3 z_1+4)-6 z_2^2}{(1-x_1)^2 x_2} 
\\ &+& \frac{x_1^2 \left(3 {(\Delta^{2,0}_{0,2})}^2-{\Delta^{0,0}_{2,2}}+{\bar{\Delta}^{0,2}_{2,2}}\right)+{\Delta^{0,3}_{2,1}} ({\Delta^{0,3}_{2,1}}-{\Delta^{2,0}_{0,2}}-z_2)}{(1-x_1)^2 x_1 x_2} \, ,
\eeqn
for the $C_A$ terms.

Finally, if we go to transcendental weight 2, we can write the term proportional to $D_A$ making use of
\beqn
F^{(2,D_A)}_1 &=& {\cal R}\left(x_1,x_3\right)\, ,
\\ \nn C^{(2,D_A)}_1 &=& \frac{2 \left(x_2 \left(x_3 \Delta^{0,1}_{3,2}+{\Delta^{3,0}_{0,3}} ({\Delta^{0,1}_{2,3}}+z_1)+x_2^3+2 x_2 x_3 z_1\right)+(\Delta^{0,1}_{3,2})^2\right)}{x_1 x_2^3} 
\\ &-& \frac{4 {\Delta^{1,0}_{0,1}} (x_3-z_1)}{x_1 x_2} \, ,
\eeqn
while for the component proportional to $C_A$ we require
\beqn
\nn F^{(2,C_A)}_1 &=& \frac{\pi ^2}{6}-2 \li{1-\frac{x_1}{1-z_1}}-2\li{1-\frac{z_2}{1-z_1}}+2 \li{1-z_1}
\\ &+& 2 \log (x_2) \log (1-z_1)\, + (1\leftrightarrow 2) ,
\\ F^{(2,C_A)}_2 &=& \log (x_1) \log (x_2) \, ,
\\ F^{(2,C_A)}_3 &=& \frac{\pi ^2}{4}-\li{1-x_1}-\log (x_1) \log (z_1) + (1\leftrightarrow 2) \, ,
\\ F^{(2,C_A)}_4 &=& \log \left(\frac{x_1}{1-z_1}\right) \log \left(\frac{1-z_1}{z_1 z_2}\right)-\log (x_2) \log (1-z_1) \, ,
\eeqn
with the associated coefficients
\beqn
C^{(2,C_A)}_1 &=& \frac{z_1^2}{x_1 x_2} \, ,
\\ C^{(2,C_A)}_2 &=& \frac{z_2-x_1 \Delta^{2,0}_{0,2} -x_1}{x_1 \Delta^{1,0}_{0,1}}+\frac{1-x_2}{x_1} \, ,
\eeqn
\beqn
C^{(2,C_A)}_3 &=& 2\frac{2 (1-z_1) z_1-x_1^2+3 x_1-1}{x_1 x_2}-\frac{2 (2 z_1+1)}{x_2}  \, ,
\\ \nn C^{(2,C_A)}_4 &=& \frac{((1-z_2)^2+ z_2^2)\Delta^{2,2}_{1,2}}{x_1 x_2 z_3}+\frac{z_2-x_1 (\Delta^{2,0}_{0,2}+1)}{x_1 \Delta^{1,0}_{0,1}}+\frac{x_2+2 \left(2 z_2^2+z_2-1\right)}{x_1}
\\ \nn &+& \frac{z_3 (3 (x_2+2) z_2-4-x_1 (1-z_2))}{x_1 x_2}+\frac{3-2 (x_1+3) (1-z_2) z_2}{x_1 x_2}
\\ &+&\frac{(x_2+3) z_3^2}{x_1 x_2}\, .
\eeqn
It is interesting to notice that these expressions are very compact, especially if we compare them with the analogous expansion for $q \to q \gamma \gamma$ and $q \to q g \gamma$. This fact is related with the presence of a initial-state vector particle, as we discuss in the next section.

\section{Processes started by photons}
\label{sec:photons}
Finally, we also considered NLO QCD corrections to splittings functions started by photons. These objects might be relevant to describe the decay of a virtual photon into three on-shell massless particles up to NLO accuracy in QCD. Following the discussion presented in the previous sections, we consider the two photon-started triple splitting processes: 
$\gamma \to q \bar{q} \gamma$ and $\gamma \to q \bar{q} g$. Moreover, let's recall that $\gamma \to ng$ vanish at tree-level for all $n \in \mathbb{N}$ due to decoupling identities, which implies that $\gamma \to ng$ must be finite beyond LO.

Besides presenting the results, we take advantage of their simplicity in order to extract some interesting conclusions about their functional structure.

\subsection{$\gamma \to q \bar{q} \gamma$}
Beginning with the LO splitting amplitude, we have
\beqn
\Spamplitude^{(0)(a_1,a_2)}_{\gamma \to q_1 \bar{q}_2 \gamma_3} &=& \frac{e^2_q g^2_e \mu^{2\ep} \IDC_{a_1 a_2}}{ {s_{123}}} \, \bar{u}(p_1) 
\left( \frac{\slashed{\ep}(p_3) \slashed{p}_{13} \slashed{\ep}(\wp)}{s_{13}}-\frac{\slashed{\ep}(\wp) \slashed{p}_{23} \slashed{\ep}(p_3)}{s_{23}} \right) 
  \, v(p_2) \, \,
\, ,
\label{SplittingLOA-qqbA}
\eeqn
where it is possible to appreciate that this amplitude is just the color stripped contribution to $\sp^{(0)}_{q_1 \bar{q}_2 \gamma_3}$. Explicitly,
\beqn
\Spamplitude^{(0)(a_1,a_2;\alpha)}_{q_1 \bar{q}_2 \gamma_3} &=& \frac{\g}{g_e e_q} \SUNT^{\alpha}_{a_1 a_2} \, \Spamplitude^{(0)(a_1,a_2)}_{\gamma \to q_1 \bar{q}_2 \gamma_3} \, ,
\eeqn
which trivially implies that the unpolarized splitting function is given by
\beqn
\la \hat{P}_{\gamma \to q_1 \bar{q}_2 \gamma_3}^{(0)} \ra&=& \frac{C_A g^2_e e^2_q}{C_F \g^2} \la \hat{P}_{q_1 \bar{q}_2 \gamma_3}^{(0)} \ra ,
\eeqn
using results from the previous section. Note that symmetry under the exchange $1 \leftrightarrow 2$ is obviously inherited. 

\begin{figure}[htb]
	\centering
	\includegraphics[width=0.80\textwidth]{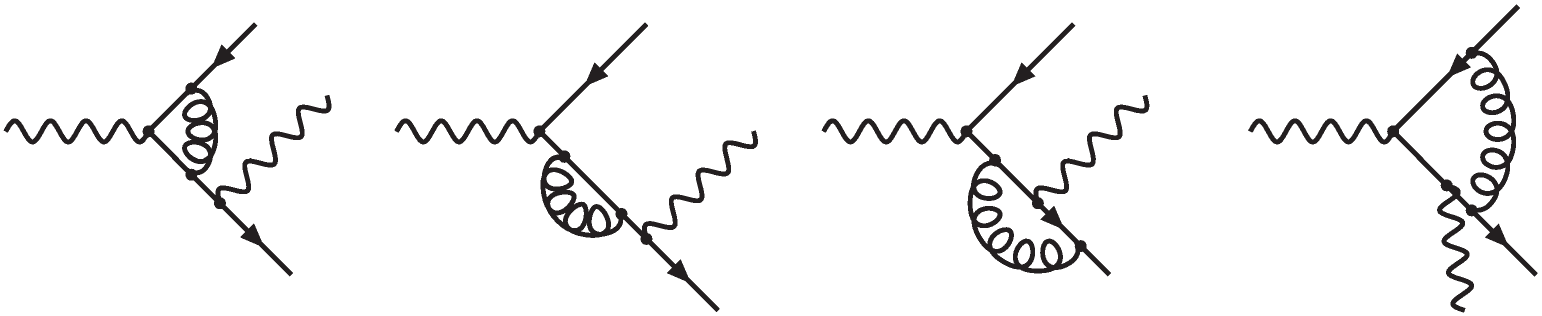}
	\caption{Representative diagrams contributing to the $\gamma \to q \bar{q}  \gamma$ splitting amplitude. We neglect self-energy corrections to the parent photon because they contribute at a higher QED order.}
	\label{fig:Diagramas_A-qqbA}
\end{figure}

When we considered $g \to q \bar{q} \gamma$ in Section \ref{sec:partons}, we arranged the different contributions to the NLO splitting function in order to identify Abelian and non-Abelian terms. We saw that purely Abelian terms were given by compact expressions which can be written in terms of logarithms and just one weight 2 function. Since $\gamma \to q \bar{q} \gamma$ is proportional to the Abelian part, its splitting function is extremely simple. In Fig. \ref{fig:Diagramas_A-qqbA} we show all the Feynman diagrams required in the computation. Notice that self-energy corrections to the incoming photon are not taken into account because they contribute to higher-orders in QED.

Following the same recipe for the other processes, we start studying the IR divergent structure. According to Catani's formula, we have
\beqn
I^{(1)}_{\gamma \to q_1 \bar{q}_2 \gamma_3} (p_1, p_2, p_3; \wp) &=& \frac{\CG \g^2}{\ep^2} \left(\frac{-s_{123}-\imath 0}{\mu^2}\right)^{-\ep } \left[-2 C_F x_3^{-\ep} - 2 \ep \gamma_q\right] \, ,
\label{eq:ICqqbargamma}
\eeqn
and we found a complete agreement with our results. On the other hand, after performing the full computation we notice that the NLO splitting function can be expressed as
\beqn
\la \hat{P}_{\gamma \to q_1 \bar{q}_2 \gamma_3}^{(1)\,{\rm fin.}} \ra&=& 2 C_F \, c^{\gamma \to q \bar{q} \gamma} \, \left[ \la \hat{P}_{q_1 \bar{q}_2 \gamma_3}^{(1,D_A)\,{\rm fin.}} \ra +  \, (1 \leftrightarrow 2)\,\right]\, ,
\eeqn
as it was expected based in a naive Feynman diagram comparison between these processes. Here we introduced the global prefactor
\beqn
c^{\gamma \to q \bar{q} \gamma} &=& \frac{C_A e^4_q g^4_e \g^2}{2(1-\ep)} \, ,
\eeqn
in order to simplify the notation.

To conclude this section, let's make a comment about crossing identities for splitting functions (and amplitudes). In Section \ref{sec:partons}, we presented $q \to q \gamma \gamma$ and the simplified weight 2 contribution to the NLO correction could be expressed in terms of 11 functions. However $\la \hat{P}^{(1)\,{\rm fin.}}_{\gamma \to q_1 \bar{q}_2 \gamma_3} \ra \left. \right|_{w=2}$ is proportional to just one of these functions. Since crossing relations are (rational) variable transformations, they can not modify the dimension of the space of functions which span the transcendental weight 2 contribution. We can explain this situation reasoning as follows. To obtain $\gamma \to q \bar{q} \gamma$ we need to perform the exchange $2 \leftrightarrow P$. This implies changing the parent parton, which is a \textit{distinguished} particle from a kinematical point of view (i.e. it is slightly off-shell and this fact involves that self-energy corrections can not be neglected). To be more explicit, in $q \to q \gamma \gamma$ the parent quark is off-shell and outgoing photons are on-shell, but in $\gamma \to q \bar{q} \gamma$ we have one off-shell photon and the quark that has been moved to the final state loses its virtuality. So, strictly speaking, kinematics forces us to consider that the particle content of these splittings is not the same.

\subsection{$\gamma \to q \bar{q} g$}
\begin{figure}[htb]
	\centering
		\includegraphics[width=0.66\textwidth]{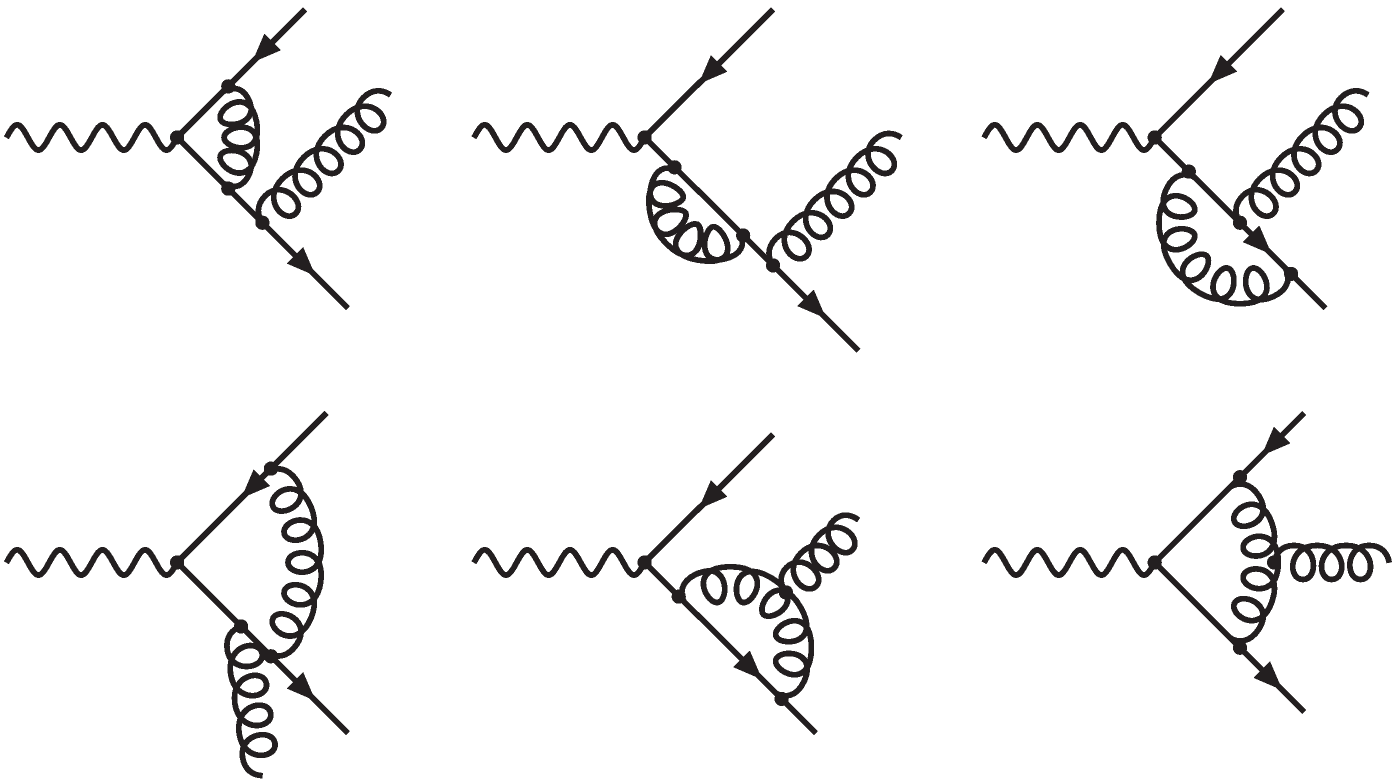}
	\caption{Representative diagrams contributing to the $\gamma \to q \bar{q}  g$ splitting amplitude.}
	\label{fig:Diagramas_A-qqbg}
\end{figure}

Finally, we arrive to the last non-trivial triple splitting process with photons: $\gamma \to q \bar{q} g$. The corresponding Feynman diagrams at NLO are shown in Fig. \ref{fig:Diagramas_A-qqbg}.

The LO splitting amplitude can be written as
\beqn
\nn \Spamplitude^{(0)(a_1,a_2,\alpha_3)}_{\gamma \to q_1 \bar{q}_2 g_3} &=& \frac{e_q g_e \g \mu^{2\ep} \SUNT^{\alpha_3}_{a_1 a_2}}{ {s_{123}}} \, \bar{u}(p_1) 
\left( \frac{\slashed{\ep}(p_3) \slashed{p}_{13} \slashed{\ep}(\wp)}{s_{13}}-\frac{\slashed{\ep}(\wp) \slashed{p}_{23} \slashed{\ep}(p_3)}{s_{23}} \right) 
  \, v(p_2) \,
\\ &=& \frac{\g}{g_e e_q} \SUNT^{\alpha_3}_{a_1 a_2} \, \Spamplitude^{(0)(a_1,a_2)}_{\gamma \to q_1 \bar{q}_2 \gamma_3}	\, ,
\label{SplittingLOA-qqbg}
\eeqn
where we used previous results to obtain the last line. In fact, using that relation with the $\gamma \to q \bar{q} \gamma$ splitting, the corresponding splitting function at LO turns out to be
\beqn
\la \hat{P}_{\gamma \to q_1 \bar{q}_2 g_3}^{(0)} \ra &=& \frac{\g^2 C_F}{g_e^2 e_q^2} \la \hat{P}_{\gamma \to q_1 \bar{q}_2 \gamma_3}^{(0)} \ra = 2 C_A C_F \, \la \hat{P}_{q_1 \bar{q}_2 \gamma_3}^{(0)} \ra \, \, ,
\eeqn
which can easily be related with different splitting functions interchanging photons and gluons. Note that we are not changing kinematics but only the color structure.

In order to express results for the NLO correction, we follow the same strategy used with $g \to q \bar{q} \gamma$. Since there are contributions proportional to both $C_A$ and $D_A=2C_F-C_A$, we treat them separately. The contribution proportional to $D_A$ is just the one coming from $\la \hat{P}_{\gamma \to q_1 \bar{q}_2 \gamma_3}^{(1)} \ra$ (with a factor 2 of difference due to the definition of $D_A$). To check the global IR divergent structure, we extract the $\ep$
poles from the splitting function and compare them with the expression provided by Catani's formula, i.e.
\beqn
\nn I^{(1)}_{\gamma \to q_1 \bar{q}_2 g_3} (p_1, p_2, p_3; \wp) &=& 
I^{(1)}_{\gamma \to q_1 \bar{q}_2 \gamma_3} (p_1, p_2, p_3; \wp) \nn \\ 
&+& \frac{\CG \g^2 C_A}{\ep^2} \left(\frac{-s_{123}-\imath 0}{\mu^2}\right)^{-\ep } \left(x_3^{-\ep}-x_1^{-\ep}-x_2^{-\ep} \right)~,
\eeqn
finding a complete agreement. Note that we used the operator 
$\IC^{(1)}_{\gamma \to q_1 \bar{q}_2 \gamma_3}$ from \Eq{eq:ICqqbargamma}
to simplify the results. Moreover, following this idea, we write the NLO 
correction to the splitting function as
\beqn
\la \hat{P}_{\gamma \to q_1 \bar{q}_2 g_3}^{(1)} \ra&=& D_A \la \hat{P}^{(1,D_A)}_{\gamma \to q_1 \bar{q}_2 \gamma_3} \ra + C_A \la \hat{P}^{(1,C_A)}_{\gamma \to q_1 \bar{q}_2 g_3} \ra \; + \; {\rm c.c.} \, ,
\eeqn
with
\beqn
\nn \la \hat{P}^{(1,D_A)}_{\gamma \to q_1 \bar{q}_2 g_3} \ra &=& \frac{1}{2 C_F} 
\left[ {\rm Re} \left( I^{(1)}_{\gamma \to q_1 \bar{q}_2 \gamma_3} (p_1,p_2,p_3;\wp) \right) \,  
\la \hat{P}_{\gamma \to q_1 \bar{q}_2 g_3}^{(0)} \ra + 
\frac{c^{\gamma \to q \bar{q} g}}{c^{\gamma \to q \bar{q} \gamma}} 
\la \hat{P}^{(1)\,{\rm fin.}}_{\gamma \to q_1 \bar{q}_2 \gamma_3} \ra \right]
\\ &=& \frac{\g^2}{2 e_q^2 g_e^2} \left[ {\rm Re}
\left( I^{(1)}_{\gamma \to q_1 \bar{q}_2 \gamma_3} (p_1,p_2,p_3;\wp) \right) \,  
\la \hat{P}_{\gamma \to q_1 \bar{q}_2 \gamma_3}^{(0)} \ra + 
\la \hat{P}^{(1)\,{\rm fin.}}_{\gamma \to q_1 \bar{q}_2 \gamma_3} \ra \right]~,
\label{EcuacionP1DAA-qqbarg1}
\eeqn
where we introduced the global prefactor
\beqn
c^{\gamma \to q \bar{q} g} &=& \frac{C_A C_F \, e_q^2 g_e^2 \g^4}{2(1-\ep)} \, .
\eeqn
To treat the finite contribution, we express it as
\beqn
\la \hat{P}_{\gamma \to q_1 \bar{q}_2 g_3}^{(1)\,{\rm fin.}} \ra&=& c^{\gamma \to q \bar{q} g}  \left[  D_A \la \hat{P}^{(1,D_A)\,{\rm fin.}}_{\gamma \to q_1 \bar{q}_2 g_3} \ra + C_A \la \hat{P}^{(1,C_A)\,{\rm fin.}}_{\gamma \to q_1 \bar{q}_2 g_3} \ra \right] \, ,
\eeqn
where the finite Abelian contribution is given by
\beqn
\la \hat{P}_{\gamma \to q_1 \bar{q}_2 g_3}^{(1,D_A)\,{\rm fin.}} \ra &=& \frac{\g^2}{2 e_q^2 g_e^2} \la \hat{P}^{(1)\,{\rm fin.}}_{\gamma \to q_1 \bar{q}_2 \gamma_3} \ra =c^{\gamma \to q \bar{q} g} \, \la \hat{P}_{q_1 \bar{q}_2 \gamma_3}^{(1,D_A)\,{\rm fin.}} \ra\, .
\eeqn
The novel contributions originated in the non-Abelian part are contained inside $\la \hat{P}^{(1,C_A)\,{\rm fin.}}_{\gamma \to q_1 \bar{q}_2 g_3} \ra$. Classifying these terms according to its transcendental weight, we get
\beqn
\la \hat{P}^{(1,C_A)\,{\rm fin.}}_{\gamma \to q_1 \bar{q}_2 g_3} \ra &=& C^{(0,C_A)} + C^{(1,C_A)} F^{(1)}  + C^{(2,C_A)} F^{(2,C_A)} + \, (1 \leftrightarrow 2) \, ,
\label{Peso2_Aqqbg}
\eeqn
where the rational coefficient is given by
\beqn
\nn C^{(0,C_A)} &=& \frac{16-7 x_2-2 z_1 z_2+(1-z_1)^2-15 z_2}{x_1}-\frac{z_1^2}{(1-x_1) x_2}-8\frac{z_1^2+(1-z_1)^2}{x_1 x_2}
\\ &+& \frac{2 z_1 (1-z_3)-x_2 (1-z_1)^2-(x_2+1) z_1}{(1-x_1) x_1}\, ,
\eeqn
and
\beqn
F^{(1)} &=& \ln{x_1} \, ,
\\ F^{(2,C_A)} &=& {\cal R}\left(x_1,x_2\right) \, ,
\eeqn
are the functions which expands the spaces of transcendentality 1 and 2, respectively. It is crucial to appreciate that $F^{(2,C_A)}$ is the same function involved in the $\gamma \to q \bar{q} \gamma$ (aside from a permutation of the kinematical variables). We will return to this point in the next subsection. Finally, in order to get the full NLO correction to this splitting function, the coefficients listed in \Eq{Peso2_Aqqbg} are given by
\beqn
\nn C^{(1,C_A)} &=& \frac{{z_2} ({x_2} (4 {x_1} {z_1}+{x_1}-1)+2 {x_3} {z_1})+{x_2} ({x_1} (({x_2}-1) {z_1}+{x_2}-3)-2 {x_2}+3)}{({x_1}-1)^2 {x_1} {x_2}}
\\ &+& \frac{3 {x_2}^2+5 {x_2} ({z_2}-1)+3 {z_2}^2-4 {z_2}+1}{{x_1} {x_2}} -\frac{(1-{x_2})^2 {z_1}^2}{(1-{x_1})^2 {x_1} {x_2}}  \, ,
\\ C^{(2,C_A)} &=& \KerLOA \, .
\eeqn
Again, it is possible to appreciate that this splitting can not easily be related to $g \to q \bar{q} \gamma$. In particular, in Section \ref{sec:partons} we showed that 7 functions were required to expand the component of transcendental weight 2 for $\la \hat{P}_{q_1 \bar{q}_2 \gamma_3}^{(1)\,{\rm fin.}} \ra$. However, $\la \hat{P}_{\gamma \to q_1 \bar{q}_2 g_3}^{(1)\,{\rm fin.}} \ra$ involves a $2$-dimensional transcendental weight 2 space.

\subsection{Further discussions on splittings started by photons}
Since the results shown in this section are very simple (compared with those presented in Section \ref{sec:partons}), we can go further and discuss about the structure of triple splittings started by photons.

Let's start with $\gamma \to q \bar{q} \gamma$. If we perform a direct computation without making an $\ep$-expansion, the NLO contribution to the splitting function can be expressed as
\beqn
\la \hat{P}_{\gamma \to q_1 \bar{q}_2 \gamma_3}^{(1)} \ra &=&  A_1^{(4)}\left(z_j,x_k;\ep\right) I_1^{(4)} + \sum_{i=1}^{3} A^{(2)}_{i}\left(z_j,x_k;\ep\right) I_i^{(2)} \, + \left(1 \leftrightarrow 2\right) \, \, ,
\eeqn
where coupling constants and color factors are absorbed inside the coefficients $A$. Here $A_{i}^{(4)}$ and $A_{i}^{(2)}$ are rational functions of the kinematical variables $\left\{x_i,z_j\right\}$ and $\ep$. Branch-cuts are defined by the Feynman integrals $I_1^{(4)}$ (box) and $I^{(2)}_{i}$ (bubbles). Moreover, using $d=4-2\ep$, the explicit expression for the box integral is
\beqn
\nn I^{(4)}_1 &=&\int_q \, \frac{\mu^{2\ep}}{q^2 (q+p_{1})^2 (q-p_{2})^2 (q-p_{23})^2} = \frac{2\CG}{\ep^2 \, x_1 x_3 s_{123}^2} \left(\frac{-s_{123}-\imath 0}{\mu^{2}}\right)^{-\ep} 
\\ \nn &\times& \left[x_1^{-\ep} \, _2 F_1 \left(1,-\ep;1-\ep;-\frac{x_2}{x_3}\right) + x_3^{-\ep} \, _2 F_1 \left(1,-\ep;1-\ep;-\frac{x_2}{x_1}\right)  \right.
\\ &-& \left. \, _2 F_1 \left(1,-\ep;1-\ep;-\frac{x_2}{x_1 x_3}\right)\right] \, ,
\eeqn
while bubbles are given by
\beqn
I^{(2)}_1 &=& \int_q\, \frac{\mu^{2\ep}}{q^2 (q-p_{123})^2} =\frac{\CG}{\ep(1-2\ep)} \left(\frac{-s_{123}-\imath 0}{\mu^{2}}\right)^{-\ep} \, ,
\\ I^{(2)}_2 &=& \int_q \, \frac{\mu^{2\ep}}{q^2 (q-p_{23})^2} =\frac{\CG x_1^{-\ep}}{\ep(1-2\ep)} \left(\frac{-s_{123}-\imath 0}{\mu^{2}}\right)^{-\ep} \, ,
\\ I^{(2)}_3 &=& \int_q\, \frac{\mu^{2\ep}}{q^2 (q-p_{12})^2} =\frac{\CG x_3^{-\ep}}{\ep(1-2\ep)} \left(\frac{-s_{123}-\imath 0}{\mu^{2}}\right)^{-\ep} \, ,
\eeqn
where we defined
\beq
\int_q = -\imath \int \frac{d^d q}{(2 \pi)^d} \, ,
\eeq
in order to simplify the notation. We can appreciate that these integrals are known to all orders in $\ep$ which implies that the same can be said about $\la \hat{P}_{\gamma \to q_1 \bar{q}_2 \gamma_3}^{(1)} \ra$.

If we study the integrals involved in NLO corrections to $\gamma \to q \bar{q} g$, we realize that an analogous situation takes place. In other words, it turns out to be possible to write
\beqn
\nn \la \hat{P}_{\gamma \to q_1 \bar{q}_2 g_3}^{(1)} \ra &=&  A_1^{(4)}\left(z_j,x_k;\ep\right) I_1^{(4)} + A_2^{(4)}\left(z_j,x_k;\ep\right) I_2^{(4)} 
\\ &+&  \sum_{i=1}^{3} A^{(2)}_{i}\left(z_j,x_k;\ep\right) I_i^{(2)} \, + \left(1 \leftrightarrow 2\right) \, ,
\eeqn
where
\beqn
\nn I^{(4)}_2 &=& \int_q \, \frac{\mu^{2\ep}}{q^2 (q+p_{2})^2 (q-p_{3})^2 (q-p_{13})^2} = \frac{2\CG}{\ep^2 \, x_1 x_2 s_{123}^2} \left(\frac{-s_{123}-\imath 0}{\mu^{2}}\right)^{-\ep} 
\\ \nn &\times& \left[x_2^{-\ep} \, _2 F_1 \left(1,-\ep;1-\ep;-\frac{x_3}{x_1}\right)  + x_1^{-\ep} \, _2 F_1 \left(1,-\ep;1-\ep;-\frac{x_3}{x_2}\right) \right.
\\ &-& \left. \, _2 F_1 \left(1,-\ep;1-\ep;-\frac{x_3}{x_1 x_2}\right)\right] = \left. I^{(4)}_1 \right|_{2 \leftrightarrow 3}\, ,
\eeqn
is just a standard one-mass box integral. The corresponding coefficients can be computed to all orders in $\ep$, but their explicit expressions are a bit cumbersome.

After motivating the results for $\la \hat{P}_{\gamma \to q_1 \bar{q}_2 \gamma_3}^{(1)} \ra$ and $\la \hat{P}_{\gamma \to q_1 \bar{q}_2 g_3}^{(1)} \ra$, it is interesting to appreciate that
\beq
I^{(4)}_1 = \frac{2\CG}{x_1 x_3 s_{123}^2} 
\left(\frac{-s_{123}-\imath 0}{\mu^{2}}\right)^{-\ep} 
\left[\frac{1}{\ep^2}-\frac{\log(x_1 x_3)}{\ep} +
\frac{\log^2\left(x_1\right)+\log^2\left(x_3\right)}{2} 
- {\cal R}\left(x_1,x_3\right) \right]~,
\eeq
which explains the origin of the function ${\cal R}$. In other words, it originates from the hypergeometric functions involved in the \textit{standard} scalar boxes (extracting squared logarithms introduced by \textit{standard} bubbles). But the important fact is that this one appears in all triple-collinear splitting functions with photons and also in the antisymmetric part of $\la \hat{P}_{q_1 \bar{Q}_2 Q_3}^{(1)} \ra$ (computed in Ref. \cite{Catani:2003vu}). So, the presence of this function is due to purely kinematical reasons. Another interesting fact is that to span triple collinear splitting functions started by photons, we only used boxes (maximal topological complexity) and bubbles (minimal non-trivial topology): triangles do not contribute to this computation.

There is a last point which deserves to be discussed. We have seen that splittings started by gluons are simpler than quark-started ones. In same sense this sounds reasonable because spinors only satisfy Dirac's equation, while physical gluons should verify transversality and gauge invariance. Explicitly, when we compute splitting amplitudes we start with an off-shell amputated amplitude ${\cal A}_{\rm amp}$ and then we project it over an on-shell spinor or polarization vector, $u(\wp)$ or $\ep(\wp)$ respectively. But spinors satisfy
\beq
\slashed{\wp} u(\wp) =0 \, ,
\eeq
while on-shell massless polarization vectors verify
\beqn
\wp \cdot \ep(\wp)=0 \, ,
\\ n \cdot \ep(\wp)=0 \, ,
\eeqn
where $n$ is a null-vector which defines the light-cone gauge. On the other hand, we can write 
\beqn
\sp_{q \to a_1 \ldots a_m} &=& \frac{1}{s_{123}}\, {\cal A}^{(1,i)}_{({\rm amp},q)} \left. \right|_{\rm LCG} u_i(\wp) \, ,
\label{SplittingGENERALquarkstarted}
\\ \sp _{g \to a_1 \ldots a_m} &=& \frac{1}{s_{123}}\, {\cal A}^{(1,\mu)}_{({\rm amp},g)} \left. \right|_{\rm LCG} \ep_{\mu}(\wp) \, ,
\label{SplittingGENERALgluonstarted}
\eeqn
for quark and gluon-started splitting amplitude, respectively. So, roughly speaking, we are imposing 2 restrictions on ${\cal A}^{\mu}_{({\rm amp},g)}$ and just one on ${\cal A}^{i}_{({\rm amp},q)}$. 

We would like to understand also which is the difference between gluon and photon-started splitting functions. The last ones are extremely compact and, moreover, they do not involve integrals with LCG propagators. If we compare the Feynman diagrams required for $\gamma \to q \bar{q} \gamma$ and $g \to q \bar{q} \gamma$, we can see that non-Abelian interactions vanish completely in the first case. Moreover, changing the coupling constant and removing the global color factor, NLO QCD corrections to $\gamma \to q \bar{q} \gamma$ are exactly the same that the corresponding QED NLO corrections. Since QED is an Abelian theory, we can use a covariant gauge to compute virtual corrections and the result remains unchanged. However the same explanation can not be directly applied to $\gamma \to q \bar{q} g$ because of NA diagrams, although the central idea is related with gauge invariance. Note that splitting functions are computed using a fixed physical-gauge (LCG in particular) in order to simplify factorization properties in the collinear limit. If we restrict the analysis to the computational procedure, we find out that we are computing NLO QCD corrections to an amputated scattering amplitude with the incoming particle slightly \textit{off-shell}, and then making a projection over an \textit{on-shell} spinor/polarization vector. When the process is started by a QCD parton, there is a color flux across the off-shell parent leg. If the parent parton is a photon then
\beq
\sum_{i=1}^{\bar{m}} \SUNT^i = 0 \, ,
\eeq
because photons are color singlets. This implies that we can attach the amputated amplitude to a pair of colorless fermions (for instance, an electron-positron pair) and reconstruct the full NLO QCD correction to the \textit{on-shell} physical scattering amplitude. Since physical scattering amplitudes are gauge invariant, we can use covariant gauge and only \textit{standard} scalar integrals are required in the computation. In other words, we can write
\beqn
\nn {\cal A}^{(1)}_{e^{-} e^{+} \to a_1 a_2 a_3}\left(k_1,k_2;p_1,p_2,p_3\right)\left. \right|_{\rm LCG} &=& \bar{v}(k_2)\left(-\imath g_e \gamma^{\nu}\right)u(k_1) \, \frac{-\imath\, \eta_{\mu \nu}}{s_{123}} \, {\cal A}^{(1,\mu)}_{({\rm amp},\gamma)} \left. \right|_{\rm LCG}
\\ &=& -g_e \, C_{\rm pol} \, \left[\frac{1}{s_{123}} \, \frac{\bar{v}(k_2)\gamma_{\mu}u(k_1)}{C_{\rm pol}} \,{\cal A}^{(1,\mu)}_{({\rm amp},\gamma)} \left. \right|_{\rm LCG}\right] \, , \
\label{EcuacionPRUEBA1}
\eeqn
where we are using physical momenta for the process $e^- e^+ \to a_1 a_2 a_3$ ($a_i$ could be a QCD parton), which implies that $k_i^2=0$ (massless fermions on-shell) and $k_1+k_2=p_1+p_2+p_3$ (vector equation). Here $C_{\rm pol}$ is an arbitrary normalization factor which only depends on the kinematics. Due to gauge invariance, 
\beq
{\cal A}^{(1)}_{e^- e^+ \to a_1 a_2 a_3}\left(k_1,k_2;p_1,p_2,p_3\right) = {\cal A}^{(1)}_{e^- e^+ \to a_1 a_2 a_3}\left(k_1,k_2;p_1,p_2,p_3\right) \left. \right|_{\rm LCG} \, ,
\eeq
which implies 
\beqn
\frac{\bar{v}(k_2)\gamma_{\mu}u(k_1)}{C_{\rm pol}} \,{\cal A}^{(1,\mu)}_{({\rm amp},\gamma)} \left. \right|_{\rm LCG} &=& \frac{\bar{v}(k_2)\gamma_{\mu}u(k_1)}{C_{\rm pol}} \,{\cal A}^{(1,\mu)}_{({\rm amp},\gamma)} \, .
\eeqn
This relation is true for every $k_i$ which fulfils physical conditions. In particular, we might use 
\beq
k^{\mu}_1=\wp^{\mu} \ \ , \ \ k^{\mu}_2=\frac{s_{123}}{2 \, nP} \, n^{\mu} \ \ \, ,
\eeq
or in the inverted order. However, if we restrict external polarizations to be physical and use one of the previously mentioned choices, then
\beq
\frac{\bar{v}(k_2)\gamma_{\mu}u(k_1)}{C_{\rm pol}} = \ep^{\pm}_{\mu}\left(\wp,n\right) \, ,
\eeq
where we are applying the well-known mapping of polarization vectors into spinor chains defined in the spinor-helicity formalism (for instance, see Ref. \cite{Dixon:1996wi}). So, relying in \Eq{SplittingGENERALgluonstarted} and these results, we conclude that it is possible to use covariant gauge in the computation of loop diagrams inside a photon-started splitting amplitude. And, for the triple collinear limit, this fact directly implies that we can make the replacement $d_{\mu \nu}(q,n) \to -\eta_{\mu \nu}$ inside the loop integrals associated with the Feynman diagrams presented in Figs. \ref{fig:Diagramas_A-qqbA} and \ref{fig:Diagramas_A-qqbg}. This proof can be generalized for splittings with more colored particles in the final state.

\section{Conclusions}
\label{sec:conclusions}

In this article we present the triple collinear splitting functions at NLO accuracy for processes which involve photons. Computations are performed using DREG to make IR/UV divergences explicit and we choose to work with CDR scheme in order to simplify the analytic treatment of the expressions. The results are organized in such a way that the divergent structure is exposed explicitly. Moreover, this structure completely agrees with the predicted behavior according to Catani's formula. Besides that, the classification of terms according to their transcendental weight and exchange symmetries allows us to obtain very compact results. These splitting functions are computed in the time-like region (TL) where strict collinear factorization is fulfilled \cite{Catani:2011st}.

We have considered processes started by QCD partons and also by photons. In particular, we used these results to explore the possibility of crossing-like relations among splitting functions beyond LO. Since the parent parton is off-shell, crossing-symmetry is broken and it is not possible to relate splittings with the same particle content. The situation is different for the double collinear limit at LO because it involves a single energy scale. Anyway, even for double-collinear configurations, the study of splittings started by photons reveals that it is not possible to establish such a connection in the context of higher-order corrections in QCD$+$QED.

Besides providing explicit results for all triple-splitting functions with at least one photon, we explore the simplifications that occur in photon-started processes. We showed that results can be expressed in terms of standard scalar boxes and bubbles. Having all QCD partons on-shell imposes additional constraints which force the cancellation of LCG denominators inside loops. We give a proof of this fact based in the spinor-helicity formalism and gauge invariance.

Finally, we would like to discuss about the importance of higher orders in the $\epsilon$-expansion of the triple collinear splitting functions. One of the main motivations to compute these objects is related with ${\rm N}^{k}$LO corrections to physical cross-sections. The counter-terms involved in subtraction-like methods require the convolution of splitting functions with some factors which contain $\epsilon$-poles. For example, at NLO, the typical form of the initial-state collinear counter-term is \cite{Catani:1996vz,Frixione:1995ms}
\beqn
d\sigma^{\rm cnt}_{a_1 a_2 \to X} &=& \frac{\as}{2\pi} \sum_b \int dz \, \frac{1}{\epsilon} \la \hat{P}_{a_1 \to b P(a_1,b)}^{(0)}(z,\epsilon)\ra d\sigma_{b a_2 \to X'}^{(0)} \, ,
\eeqn
which contains ${\cal O}(\epsilon^0)$ contributions arising from ${\cal O}(\epsilon)$ terms in the splitting functions. The codes that we employ to obtain the results shown in this article can be extended to compute ${\cal O}(\epsilon)$ terms (or even higher orders), although explicit analytic expressions might be extremely lengthy. Anyway, numerical results for the counter-terms could be calculated provided that we know the corresponding $\epsilon$-expansion of the involved master integrals.


  \subsection*{Acknowledgments}
We would like to thank Stefano Catani for making useful comments in an early stage of this project. Also, we would like to thanks M. V. Garzelli for the fruitful discussions about the implementation of numerical checks. This work is partially supported by UBACYT, CONICET, ANPCyT, the
Research Executive Agency (REA) of the European Union under
the Grant Agreement number PITN-GA-2010-264564 (LHCPhenoNet),
by the Spanish Government and EU ERDF funds
 (grants FPA2011-23778 and CSD2007-00042
Consolider Ingenio CPAN) and by GV (PROMETEUII/2013/007).


\end{document}